\documentclass[letterpaper,english,reprint, aps]{revtex4-1}
\usepackage[OT1]{fontenc}
\usepackage[utf8]{inputenc}

\usepackage{verbatim}

\usepackage{amsmath}
\usepackage{amsfonts}
\usepackage{amssymb}
\usepackage{tikz}
\usepackage{caption}
\usepackage[font={small,it}]{caption}
\usepackage[toc,page]{appendix}
\usepackage{natbib}

\usepackage{todonotes}
\usepackage{rotating}

\usepackage{datatool}
\usepackage{collcell}

\usepackage{subcaption}
\usepackage{tabularx}
\makeatletter

\pdfpageheight\paperheight
\pdfpagewidth\paperwidth
\usepackage{graphicx}
\graphicspath{ {Images_article/} }

\makeatother
\usepackage{babel}

\usepackage{booktabs,array}

\newcount\rowc

\makeatletter
\def\ttabular{%
\hbox\bgroup
\let\\\cr
\def\rulea{\ifnum\rowc=\@ne \hrule height 1.3pt \fi}
\def\ruleb{
\ifnum\rowc=1\hrule height 1.3pt \else
\ifnum\rowc=6\hrule height \heavyrulewidth 
   \else \hrule height \lightrulewidth\fi\fi}
\valign\bgroup
\global\rowc\@ne
\rulea
\hbox to 10em{\strut \hfill##\hfill}%
\ruleb
&&%
\global\advance\rowc\@ne
\hbox to 10em{\strut\hfill##\hfill}%
\ruleb
\cr}
\def\endttabular{%
\crcr\egroup\egroup}

\begin{document}

\preprint{APS/123-QED}

\title{Testing the causality of Hawkes processes with time reversal}

\author{Marcus Cordi}

\affiliation{Laboratoire de Mathématiques et Informatique pour les Systèmes Complexes,
CentraleSupélec, Université Paris Saclay}

\author{Damien Challet}

\affiliation{Laboratoire de Mathématiques et Informatique pour les Systèmes Complexes,
CentraleSupélec, Université Paris Saclay}

\author{Ioane Muni Toke}

\affiliation{Laboratoire de Mathématiques et Informatique pour les Systèmes Complexes,
CentraleSupélec, Université Paris Saclay}

\date{\today}
\begin{abstract}
We show that univariate and symmetric multivariate Hawkes processes are only weakly causal: the true log-likelihoods of real and reversed event time vectors are almost equal, thus parameter estimation via maximum likelihood only weakly depends on the direction of the arrow of time. In ideal (synthetic) conditions, tests of goodness of parametric fit unambiguously reject backward event times, which implies that inferring kernels from time-symmetric quantities, such as the autocovariance of the event rate, only rarely produce statistically significant fits. Finally, we find that fitting financial data with many-parameter kernels may yield significant fits for both arrows of time for the same event time vector, sometimes favouring the backward time direction. This goes to show that a significant fit of Hawkes processes to real data with flexible kernels does not imply a definite arrow of time unless one tests it.

\begin{description}
\item [{PACS~numbers}]  \textsf{PACS~numbers}.{\small \par}
\end{description}
\end{abstract}



\maketitle

\section{Introduction}
Hawkes processes (HPs hereafter) extend Poisson processes by allowing a modulation of the current event rate as a function of the past events and are thus  self-excited Poisson processes. Accordingly, they are very useful in many
fields where the occurrence of one event increases for some time the probability of another event. Examples may be found in seismology, where an earthquake typically is followed by aftershocks \cite{ogata1988statistical, gardner1974sequence, zhuang2002stochastic, marsan2008extending}, criminology, where a fight between rival gangs may trigger various criminal retaliations \cite{mohler2011self}, neurology, where the spiking activity of individual neurons may depend on the neuron's own spiking history \cite{truccolo2005point, pillow2008spatio, london2010sensitivity}, and credit risk, where the default of one company in a portfolio may lead to the default of other companies \cite{dassios2017generalized}. 

Quite remarkably, in many papers, the goodness of fit is not quantitatively assessed, but only qualitatively with Q-Q plots (which often {\em look} good), probably because HPs are assumed to be useful extensions of Poisson processes that are either totally adequate or cannot possibly describe precisely the data, which amounts to making unverified assumptions about the goodness of fits in either case. However, recent results show that parametric fits of HPs to high-frequency financial data do pass goodness of fit tests provided that a multi-timescale kernel is used and the non-stationary baseline intensity is properly accounted for for~\cite{lallouache2016limits,omi2017hawkes}. 

HPs are causal by construction. Indeed, in the univariate case, the conditional intensity of a HPs $N_t$, or equivalently its rate of events, evolves according to
\begin{equation}
\lambda(t)=\lambda_0(t)+\int\limits_{-\infty}^tK(t-s)\mathrm{d}N_s=\lambda_0(t)+\sum\limits_{t_i<t}K(t-t_i),
\label{eq:exp_kernel}
\end{equation}
where $\lambda_0(t)$ is the baseline intensity (hereafter we will assume constant baseline intensity, i.e., $\lambda_0(t)=\lambda_0$), $K(t)$ is the kernel of the process and $t_i$ the time of event $i$: $\lambda$ is defined in a causal
way from past events, hence the direction of time is well-defined. It would thus seem foolish to fit a HP to the reverted vector of events, i.e., to the events taken in the backward direction. Accordingly, 
the belief that a time series of events with an inverted arrow of time cannot possibly be mistaken for a HP is widely established (see for example Ref.\ \cite{kirchner2017estimation}). 
 
\begin{figure}
\centering
\includegraphics[width=0.9\columnwidth]{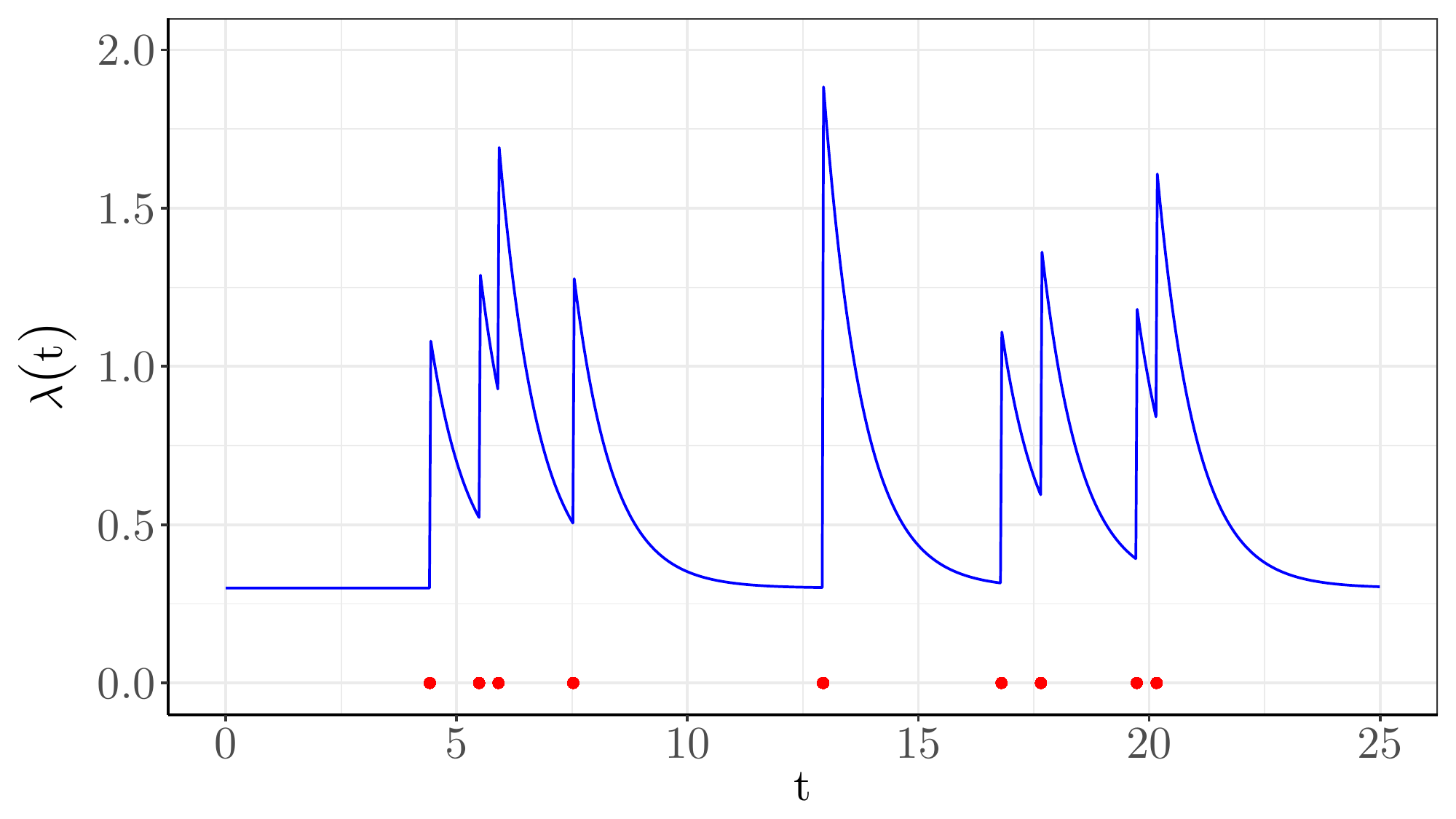}
\includegraphics[width=0.9\columnwidth]{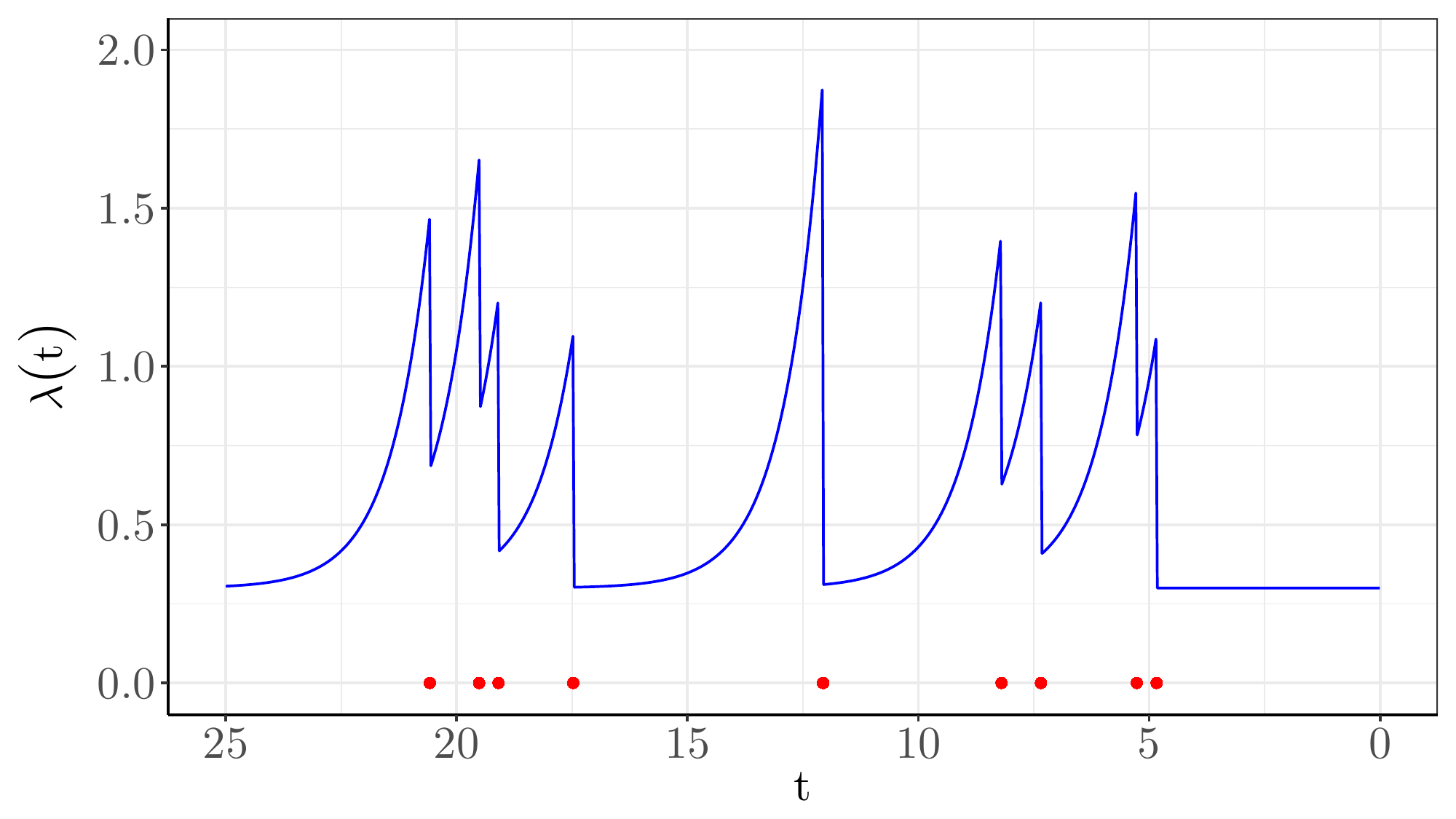}
\caption{Intensity as a function of time of a HP with an exponential kernel $K(t)=\alpha \mathrm{e}^{-\beta t}$ and  $\lambda_0=0.3$, $\alpha=0.8$ and $\beta=1.2$ for the true (top) and the time-reversed sequence of events (bottom). The red points indicate events. }
\label{fig:fig_intensity_HP_time_reversal}
\end{figure}

However, the strength of the causality of HPs is as of yet unknown and indeed the extent to which true HPs are distinctively causal depends on the method used to assess the fitted model and, when fitting them to data, on the nature of the data. As shown below, a parametric kernel estimation of univariate and symmetric multivariate HPs on synthetic data leads on {\em average} to almost the same values for both time arrows. Why this may be the case is best gathered from a classic plot that superposes the activity rate $\lambda(t)$ with the event times (Fig.\ \ref{fig:fig_intensity_HP_time_reversal}). The twist is to plot the activity rate from the same sets of events, and to let the time run backwards: the activity rate and the clustering of events are visually plausible for both directions of the time arrow.

Expectedly, goodness of fit tests are able to discriminate between a forward and a backward arrow of time for synthetic data (i.e., in an ideal setting), the latter being very often detected as not HPs. A related issue is found when one infers kernel with time-reversal symmetric quantities, which by definition yield exactly the same kernel for both arrows of time. For example, the non-parametric kernel inference of \cite{bacry2012non} is based on the covariance of event rates, which is symmetric with respect to time reversal. We show here that such kernels only rarely pass tests of goodness of fit. However, we point out that this method provides a useful approximation of the true kernel shape precisely when causality is weak (i.e. in the limit of small endogeneity), which may then help choosing a parameteric kernel family.

Fitting HPs to real data is more troublesome. For example, data collection may further degrade causality if the time resolution is too coarse.
But by far the main problem is that one does not know the shape of the kernel. We show that the more flexible the kernel, the harder it becomes for tests of goodness of fit to discriminate between the forward and backward arrows of time, sometimes yielding statistically significant fits for both time directions of the same set of events. In financial data, fits usually (and reassuringly) favour the forward arrow of time. However, there are cases when the backward arrow of time yields better fits than the forward one, at odds with the reality of financial markets, which shows that a significant fit of a weakly causal HP does not necessarily correspond to physical causality. By extension, inferring from a fit that a system is causal because of the success of a fit of a weakly causal HP should be avoided.

\section{Univariate processes}
We performed extensive numerical simulations by generating HPs with a single exponential kernel 
\begin{equation}\label{eq:HP_exp1}
K(t)=\alpha \mathrm{e}^{-\beta t}
\end{equation}
and constant baseline intensity  for a variety of parameters  with the Ogata thinning method \cite{ogata1981lewis}; results for a power-law kernel are reported in Appendix~\ref{appendix:powerlaw} and are similar to those obtained with a single exponential.

The data points are grouped according to the endogeneity of the process, defined as
\begin{equation}
n=\int\limits_{0}^{\infty}K(s)\mathrm{d}s.
\end{equation}
The endogeneity (or reflexivity as it might be referred to in the field of finance \cite{soros2003alchemy}) quantifies the level of the relative self-excitement of the process \cite{filimonov2012quantifying}. In order for the process to be stationary the endogeneity must satisfy
$n<1$. In the specific case of a HP with an exponential kernel, the endogeneity is given by $n=\frac{\alpha}{\beta}$.

Since the expected number of events of a stationary simple point process is given by
\begin{equation}
E\left[N_t\right]=\mu t,
\label{eq:exp_no_events}
\end{equation}
where $\mu=E\left[\lambda(t)\right]=\frac{\lambda_0}{1-n}$ for HPs, we have adjusted the time horizon $T$ so that all the simulations have the same expected number of events in order to allow a proper comparison between all the results obtained with different values of $n$.

In order to avoid calibration issues, we first of all remove ("burn") the non-stationary part of all simulations. The time of stationarity $t_0$ is defined as the first time the instantaneous intensity is greater or equal to the average (expected) intensity, i.e.,
\begin{equation}
t_{0}=\inf\{t\in\{t_i\}_{i=1,\dots,n}:\lambda(t)\geq\mu\}.
\end{equation}
The process is then shifted: $t_i'=t_i-t_{0}, t_i>t_0$ and $T'=T-t_0$. This requires us to modify the usual likelihood estimation, as explained below. We shall henceforth drop the prime symbols for the sake of readability.

The vector of event times obtained from the simulations (or data) correspond by definition to the forward arrow of time and will be denoted henceforth by $\{t_i^{(f)}\}_{i=1,\dots,n }$. The events in the backward arrow of time simply correspond to taking the last event of the forward time arrow as the first one in the backward arrow of time, the second last event as the second one and so on; mathematically, $t_i^{(b)}=T-t_{n+1-i}^{(f)}$. 
 
We compare the adequacy of HPs to both forward and backward event time series with three methods: the likelihood function calculated with the true parameters, Maximum Likelihood Estimation (MLE hereafter) and goodness of fit. 

\subsection{Log-likelihood}

\begin{figure}
\centering{
\includegraphics[width=0.9\columnwidth]{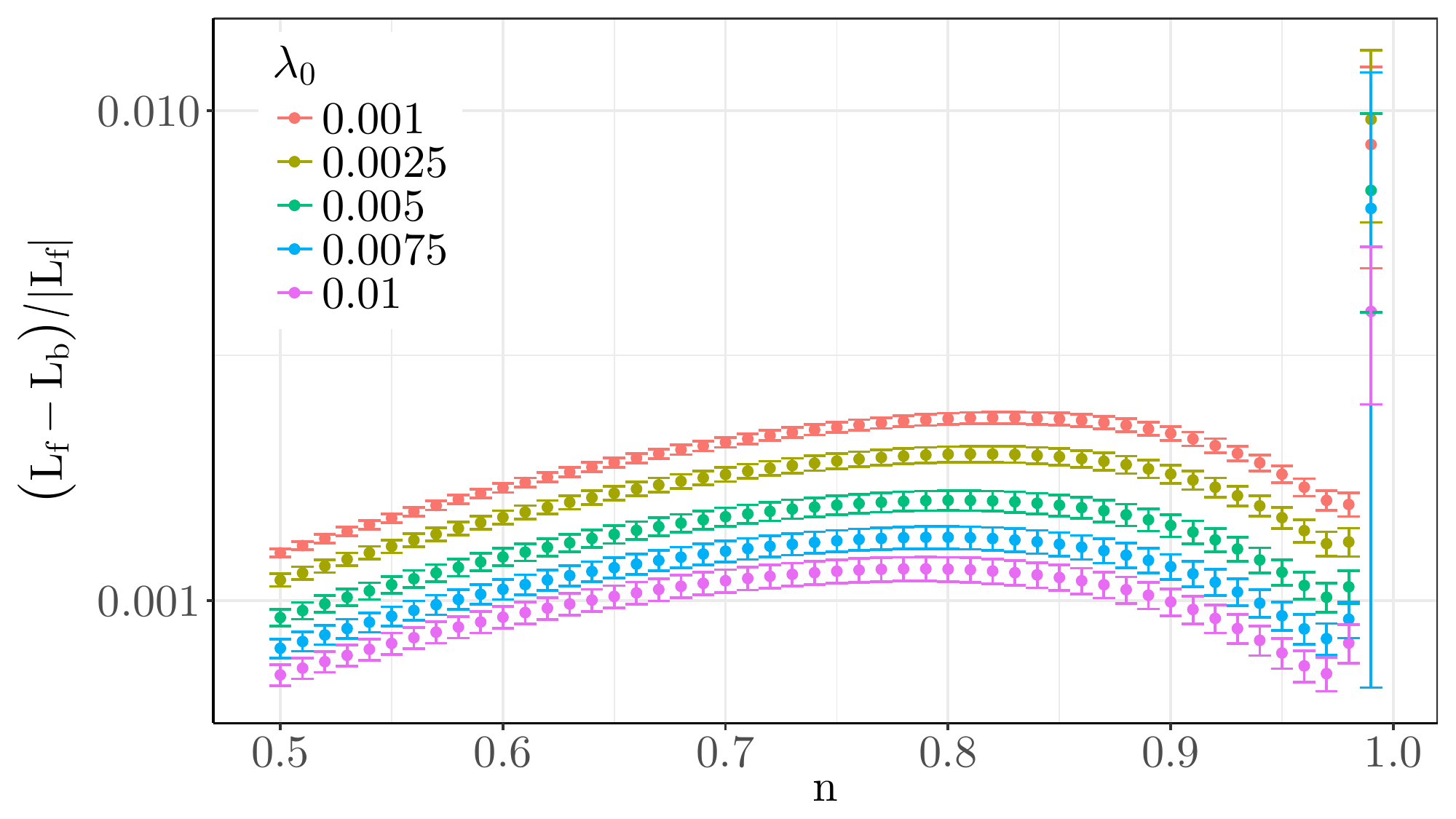}
\includegraphics[width=0.9\columnwidth]{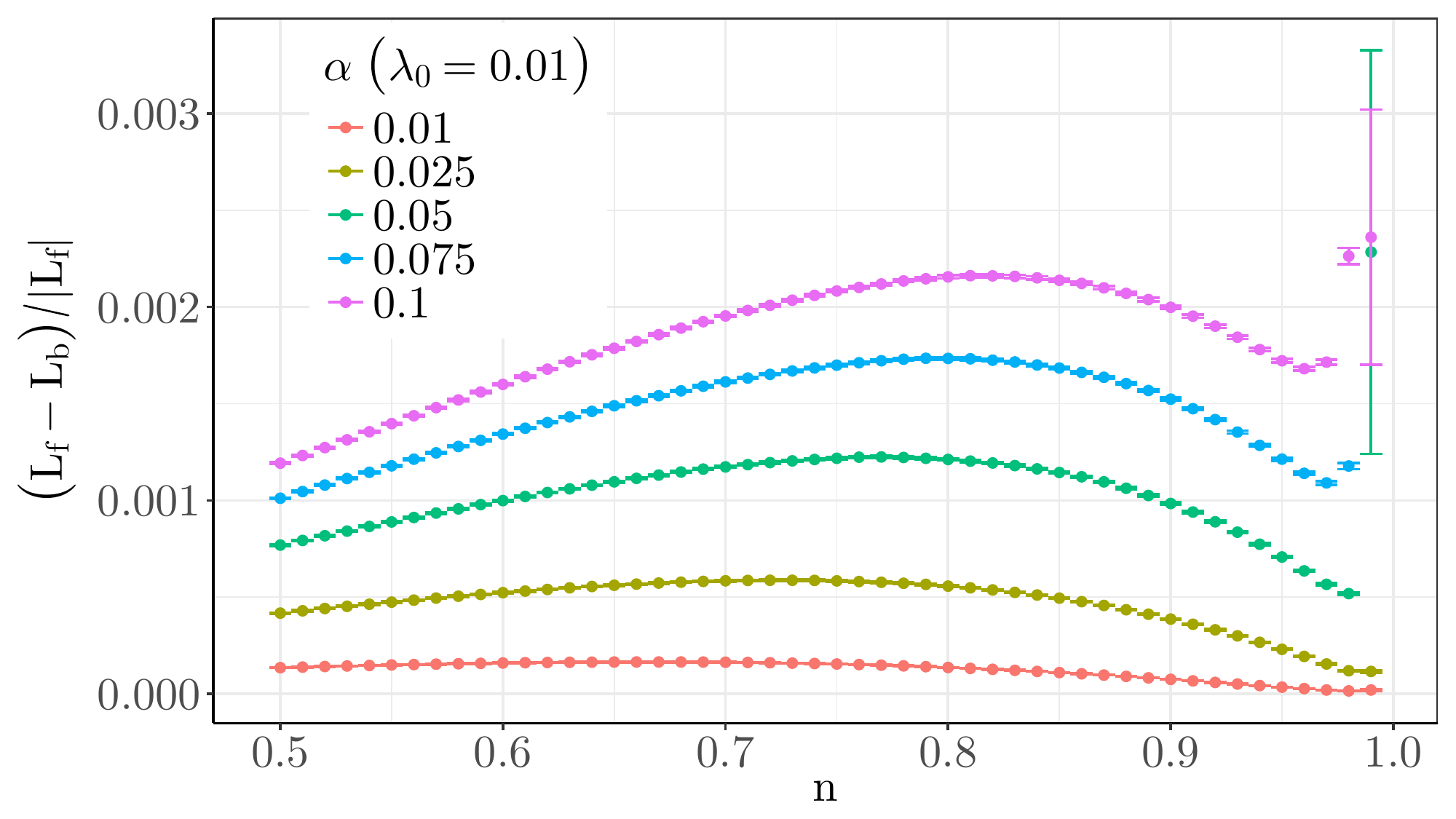}
}
\caption{Relative difference of the log-likelihood between forward and backward time arrows for a HP with an exponential kernel. All possible permutations of $\lambda_0=\{0.001, 0.0025, 0.0050, 0.0075, 0.0100\}$ and $\alpha=\{0.010, 0.025, 0.050, 0.075, 0.100\}$, with $\beta$ chosen according to the desired endogeneity $n$, are considered. The data points are grouped according to their endogeneity and averaged over 100 runs for each parameter permutation. The expected number of events is set to $10^6$.}
\label{fig:loglik_reldiff_exp}
\end{figure}

\begin{figure}
\centering{
\includegraphics[width=0.9\columnwidth]{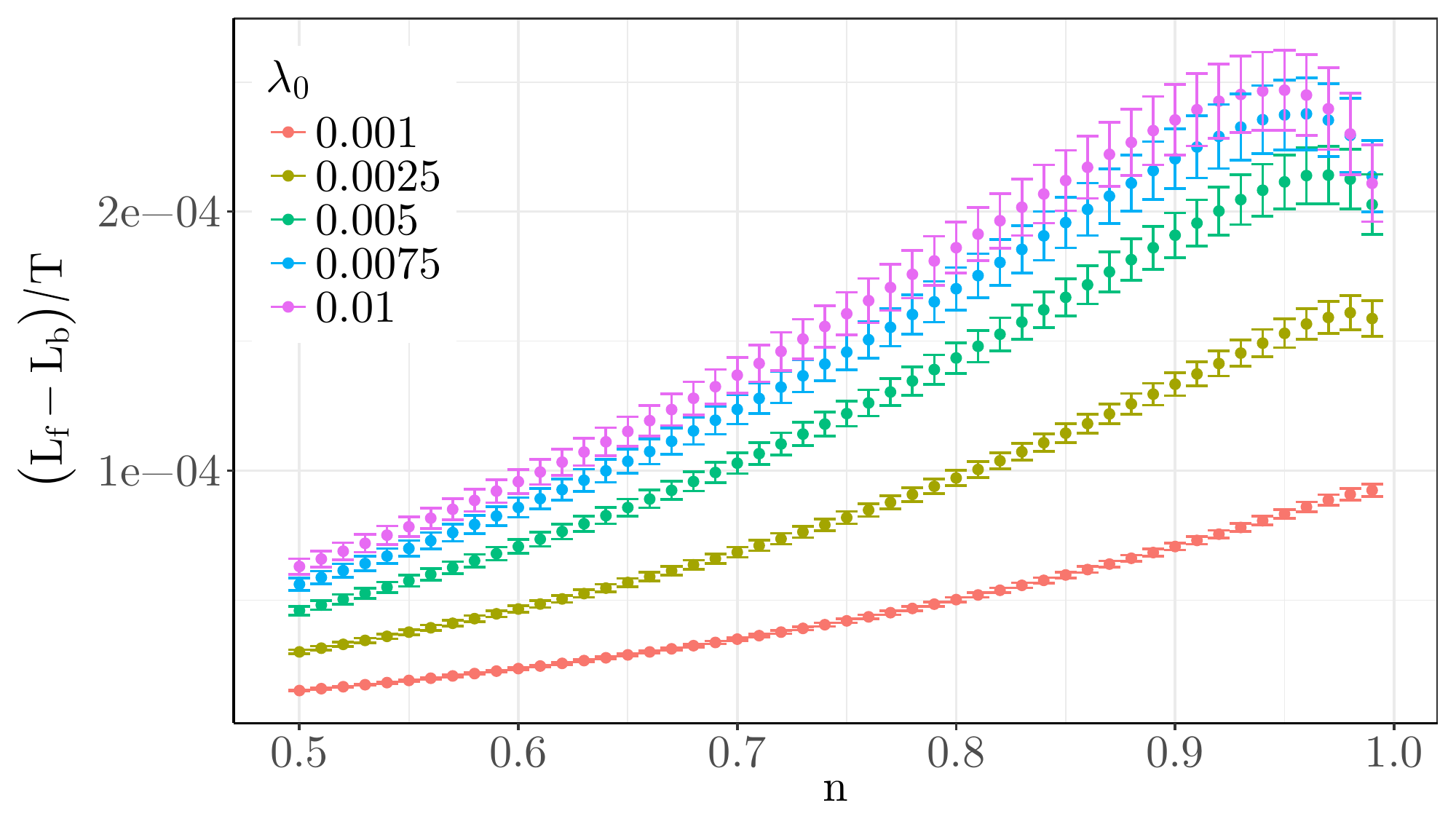}
\includegraphics[width=0.9\columnwidth]{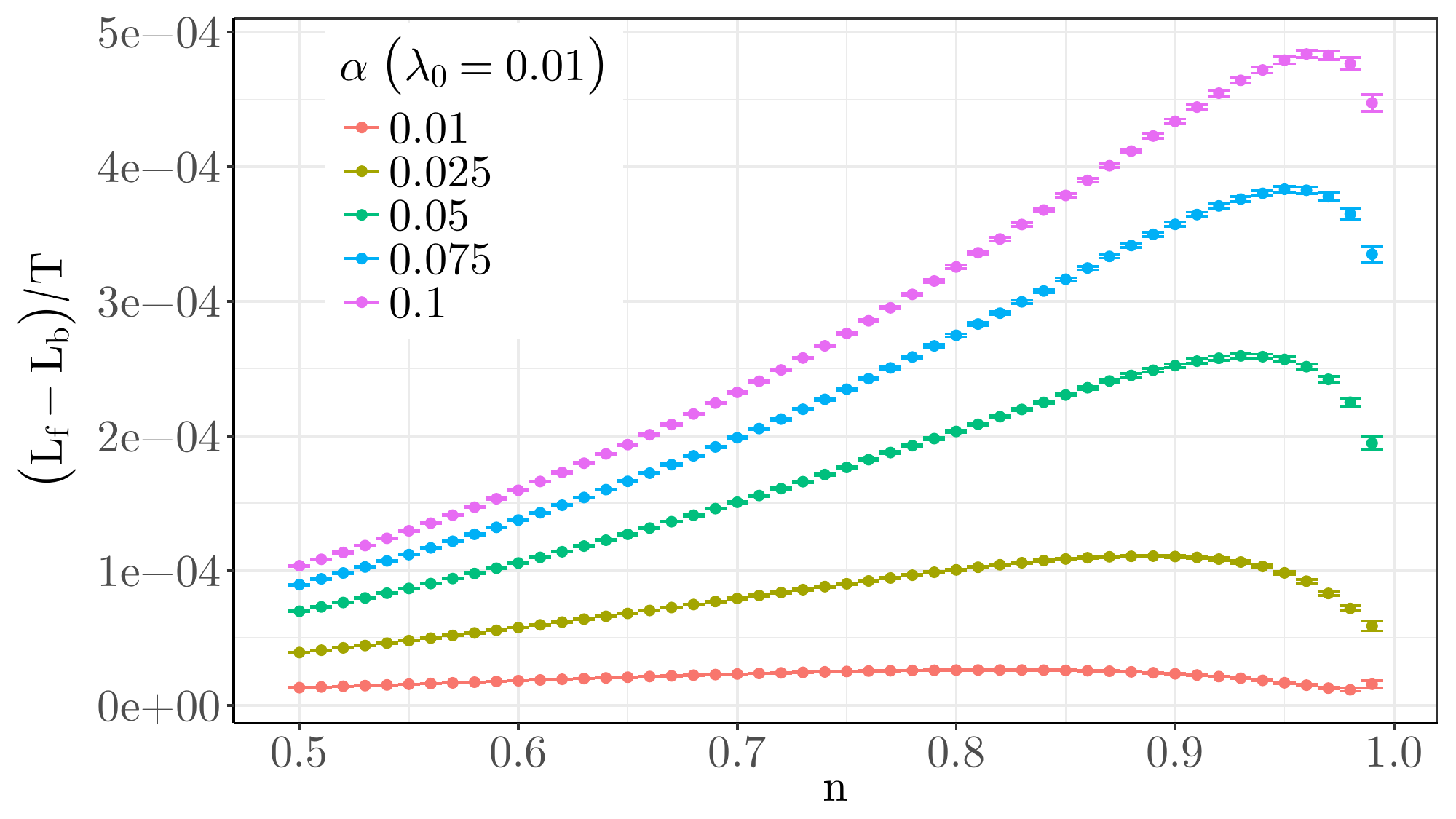}
}
\caption{Difference of the log-likelihood scaled by $T$ between forward and backward time arrows for a HP with an exponential kernel for  $\lambda_0=\{0.001, 0.0025, 0.0050, 0.0075, 0.0100\}$ and $\alpha=\{0.010, 0.025, 0.050, 0.075, 0.100\}$, while $\beta$ is adjusted to match the desired $n$. The data points are grouped according to their endogeneity and averaged over 100 runs for each parameter permutation. The expected number of events is set to $10^6$.}
\label{fig:loglik_reldiff_exp_T}
\end{figure}

The idea here is to compare the true log-likelihood, i.e., computed with the true kernel, of simulations of HPs for the real (forward) and reversed (backward) event time vectors. 
The log-likelihood of a univariate point process $N_t$ with intensity $\lambda(t)$ is written as
\begin{equation}
\ln \mathcal{L}\left(\left(N_t\right)_{t\in[0,T]}\right)=-\int\limits_0^T\lambda(s)\mathrm{d}s+\int\limits_0^T\ln\lambda(s)\mathrm{d}N_s.
\end{equation}
In the case of a HP with an exponential kernel and a constant baseline intensity, the log-likelihood is
\begin{equation}
\begin{aligned}
\ln\mathcal{L}\left(\left\{t_i\right\}_{i=1,\dots,n }\right)=-\lambda_0T-\sum\limits_{i=1}^n\frac{\alpha}{\beta}\left(1-\mathrm{e}^{-\beta(T-t_i)}\right)\\
+\sum\limits_{i=1}^n\ln\left[\lambda_0+\sum\limits_{k=1}^{i-1}\alpha\mathrm{e}^{-\beta(t_i-t_k)}\right].
\end{aligned}
\end{equation}

This expression, however, takes into account the initial non-stationary part of the process. A fair comparison between the forward and backward processes requires the removal of the non-stationary part of the process, which leads to small modifications of the above mathematical expression.

The general idea behind the modification is that if the simulation has already reached a stationary state, then the (constant) baseline intensity $\lambda_0$ should be replaced by a time-dependent baseline intensity $\lambda_0^\prime(t)$, which is given by
\begin{equation}
\lambda_0^\prime(t)=\lambda_0+\left(\frac{\lambda_0}{1-n}-\lambda_0\right)\frac{K(t)}{K(0)}.
\end{equation}
A similar procedure is developed in \cite{roueff2016locally}.
In the case of the exponential kernel we obtain
\begin{equation}
\begin{aligned}
\ln\mathcal{L}\left(\left\{t_i\right\}_{i=1,\dots,n }\right)=-\lambda_0T-\left(\frac{\lambda_0}{1-\frac{\alpha}{\beta}}-\lambda_0\right)\frac{1-\mathrm{e}^{-\beta T}}{\beta}\\
-\sum\limits_{i=1}^n\frac{\alpha}{\beta}\left(1-\mathrm{e}^{-\beta(T-t_i)}\right)\\+\sum\limits_{i=1}^n\ln\left[\lambda_0+\left(\frac{\lambda_0}{1-\frac{\alpha}{\beta}}-\lambda_0\right)\mathrm{e}^{-\beta t_i}+\sum\limits_{k=1}^{i-1}\alpha\mathrm{e}^{-\beta(t_i-t_k)}\right].
\end{aligned}
\end{equation}

In order to assess the performance of this correction we content ourselves with comparing the average difference between the MLE estimates and the true values (see Table \ref{table:diff_loglik_mod}), and see that the modified log-likelihood does indeed generally perform slightly better than the standard log-likelihood on truncated HPs.

\begin{table}
\caption{Average difference between the true parameter values and the estimations obtained via MLE for the forward (top) and backward process (bottom)) with the standard log-likelihood function (SLL) and the modified log-likelihood function (MLL) for a truncated HP (the same parameter choice as used in Figs.~\ref{fig:loglik_reldiff_exp},~\ref{fig:loglik_reldiff_exp_T} and~\ref{fig:fig_par_HP_lambda_full_par_space}, except the variable number of expected events).} 
\label{table:diff_loglik_mod}
\small
\centering
\scalebox{0.65}{
\begin{tabular}{|c|c|c|c|}
        \hline
         Forward & $\lambda_0^{(f)}$ & $\alpha^{(f)}$ & $\beta^{(f)}$ \\         
        \hline
        SLL. $E\left[N_T\right]=10^4$ & $9.901\%$ & $2.496\%$ & $2.751\%$ \\
        \hline
        MLL $E\left[N_T\right]=10^4$ & $5.497\%$ & $2.294\%$ & $2.192\%$ \\
        \hline
        SLL $E\left[N_T\right]=10^5$ & $1.205\%$ & $0.692\%$ & $0.687\%$ \\
        \hline
        MLL $E\left[N_T\right]=10^5$ & $1.056\%$ & $0.661\%$ & $0.656\%$ \\
        \hline
        SLL $E\left[N_T\right]=10^6$ & 0.332$\%$ & $0.220\%$ & $0.217\%$ \\
        \hline
        MLL $E\left[N_T\right]=10^6$ & 0.328$\%$ & $0.213\%$ & $0.210\%$ \\
        \hline
        \end{tabular}
}
\smallskip
\vspace{1cm}
\small
\centering
\scalebox{0.65}{
\begin{tabular}{|c|c|c|c|}
        \hline
        Backward & $\lambda_0^{(b)}$ & $\alpha^{(b)}$ & $\beta^{(b)}$\\         
        \hline
        SLL $E\left[N_T\right]=10^4$ & $10.476\%$ & $2.674\%$ & $2.886\%$ \\
        \hline
        MLL $E\left[N_T\right]=10^4$ & $ 6.100\%$ & $2.450\%$ & $2.461\%$ \\
        \hline
        SLL $E\left[N_T\right]=10^5$ & $1.637\%$ & $0.849\%$ & $1.020\%$ \\
        \hline
        MLL $E\left[N_T\right]=10^5$ & $1.496\%$ & $0.817\%$ & $0.988\%$ \\
        \hline
        SLL $E\left[N_T\right]=10^6$ & $1.088\%$ & $0.552\%$ & $0.797\%$ \\
        \hline
        MLL $E\left[N_T\right]=10^6$ & $1.083\%$ & $0.544\%$ & $0.790\%$ \\
        \hline
        \end{tabular}
}
\end{table}


Figure~\ref{fig:loglik_reldiff_exp} displays the average relative difference of the log-likelihood calculated with the true parameters for both time arrows. It turns out that it is surprisingly small, typically 0.2\% on average for a very large number of events, except for near-critical ($n\simeq 1$) processes. Here we see that, as one would expect, the likelihood of the forward event time series is consistently larger than that of the backward event time series. 

On average a lower baseline intensity $\lambda_0$ implies a larger difference in the log-likelihood, as one might expect since the Poissonian properties of the process are less prominent. Similarly, a larger $\alpha$  also implies a larger difference because each event carries with it a larger impact on the intensity. The difference of the forward and backward log-likelihood scaled by $T$ has a similar behaviour (see Fig.\ \ref{fig:loglik_reldiff_exp_T}).

We have checked the fraction of the simulations for which the true likelihood of the backward process is larger than that of the forward process. Expectedly, since we compute the likelihood with the true kernel, we found $8\cdot 10^{-6}$, which is to say none. One should however keep in mind that when dealing with empirical data, one faces three additional problems that may change this rosy outcome, as indeed the above situation is an ideal case. First, one does not know the true kernel shape nor its parameters. Second, the number of events in the above simulations is much larger than those of the typical dataset. Third, the question of how to deal with a non-constant baseline intensity is fundamental, but still under active investigation; the issue here is to properly discriminate between exogenous and endogenous events, i.e., to attribute time variations of the intensity to the kernel or to the baseline intensity.

\subsection{Parameter estimation}

The small difference found in the log-likelihood suggests that the estimation of the parameters based on maximum likelihood leads to fairly similar parameter values. We thus perform MLE on synthetic data; we impose that $\alpha<\beta$ in order to fulfil the requirement for a stationary process, both for the original and the time-reversed sequence of events. The few non-convergent estimations were excluded from the analysis. Since we choose as initial points the true parameter values, the optimisation is typically not required to be bound constrained and an algorithm by Nelder and Mead \cite{nelder1965simplex} is used.  When working with real-world data as in Section~\ref{sec:application_to_data}, however,  there is a need for a bound constrained optimisation and the L-BFGS-B algorithm \cite{byrd1995limited} is used.

\begin{figure}
\centering
\includegraphics[width=0.9\columnwidth]{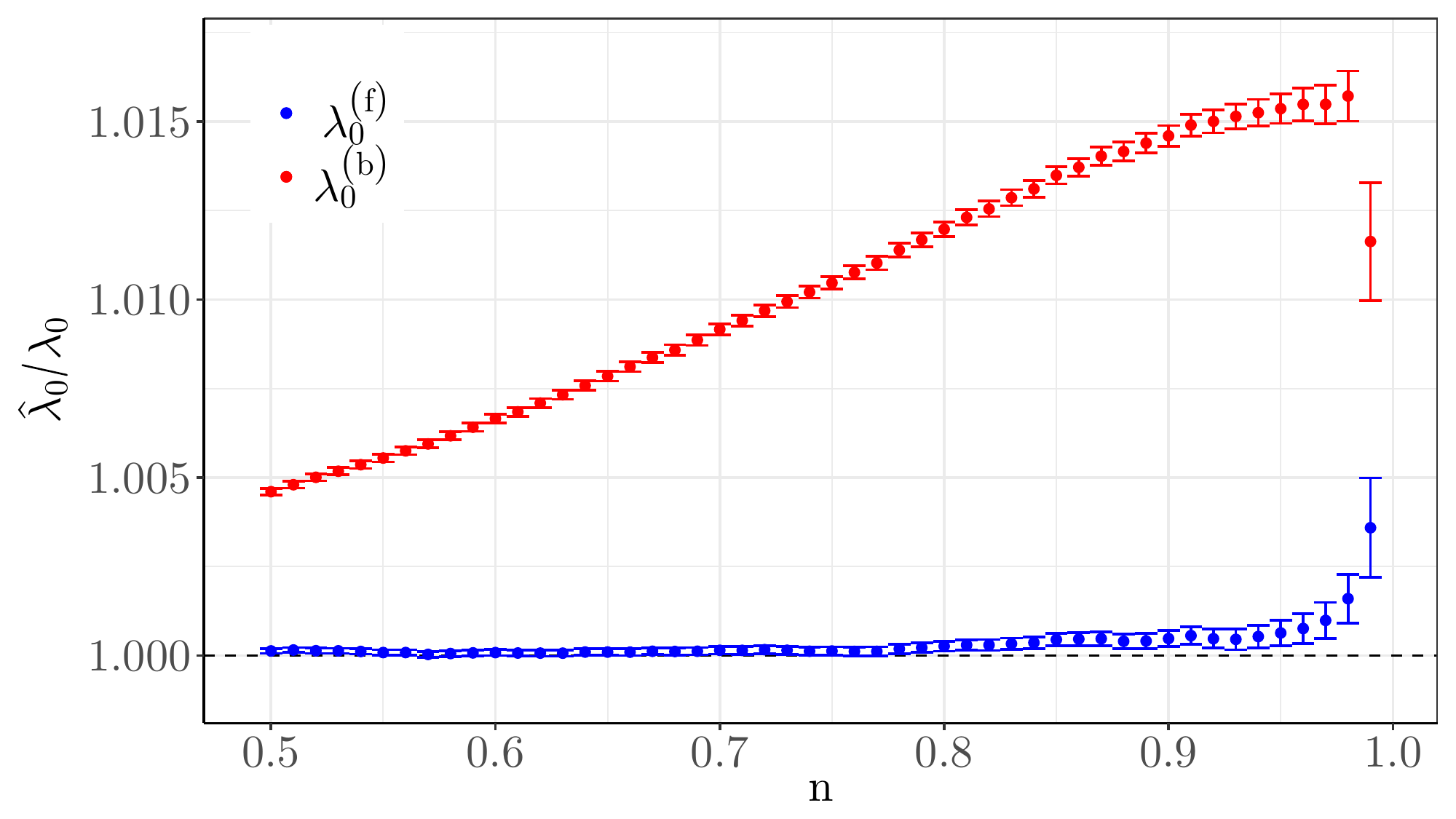}
\includegraphics[width=0.9\columnwidth]{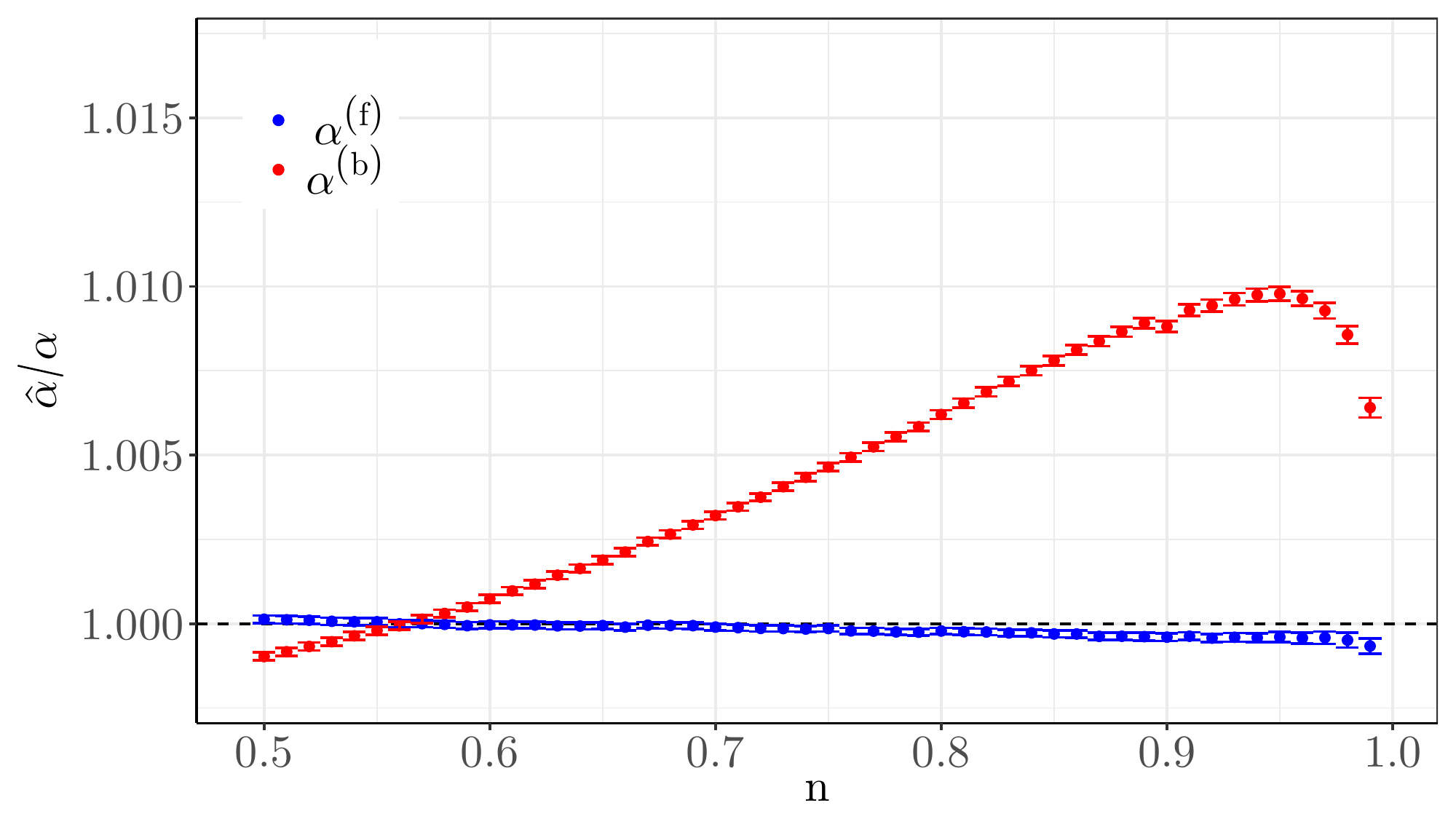}
\includegraphics[width=0.9\columnwidth]{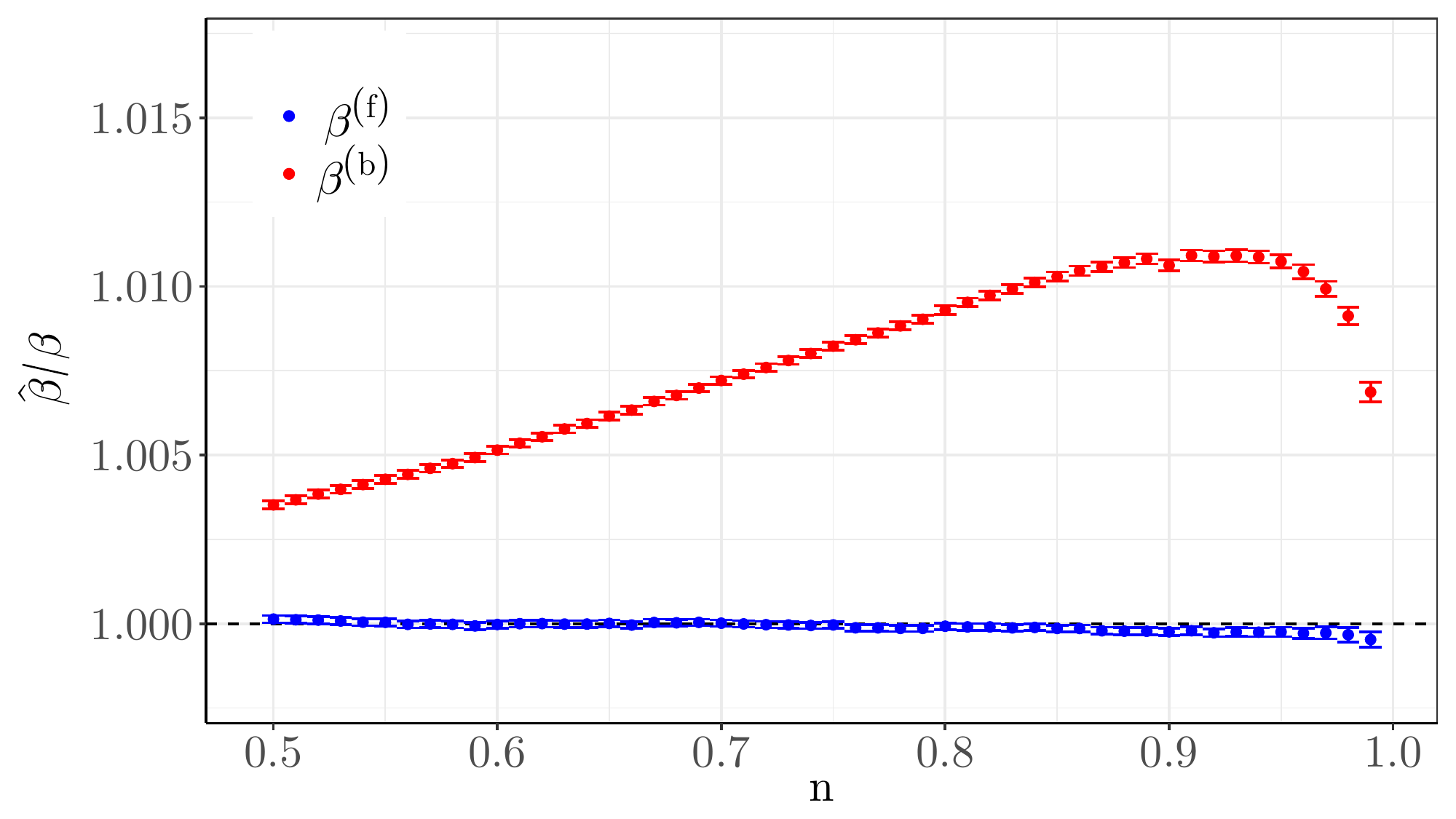}
\includegraphics[width=0.9\columnwidth]{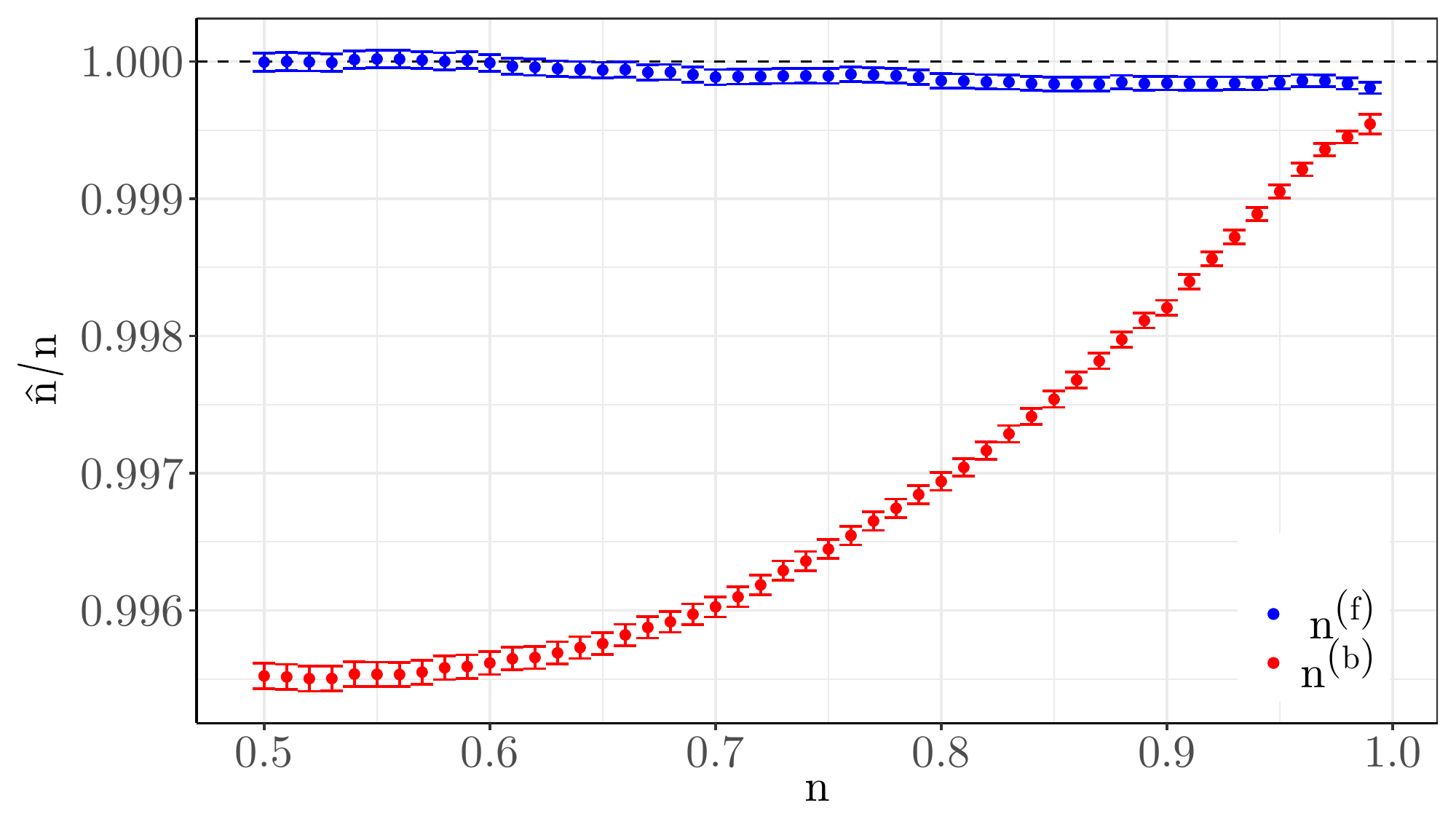}
\caption{Relative difference in the estimation of $\lambda_0$, $\alpha$, $\beta$ and $n$ in the MLE of the exponential HP for the forward (blue) and the backward process (red). All possible permutations of $\lambda_0=\{0.001, 0.0025, 0.0050, 0.0075, 0.0100\}$ and $\alpha=\{0.010, 0.025, 0.050, 0.075, 0.100\}$, with $\beta$ chosen according to the desired endogeneity $n$, are considered. The data points are grouped according to their endogeneity and averaged over 100 runs for each parameter permutation. The expected number of events is set to $10^6$.}
\label{fig:fig_par_HP_lambda_full_par_space}
\end{figure}

Unsurprisingly, Fig.~\ref{fig:fig_par_HP_lambda_full_par_space} reveals that the estimated parameters only weakly depend on the direction of the time arrow of the event time series. One notes that the baseline intensity is somewhat overestimated for the time-reversed process. One interpretation is that since causality is lost, the fitting process must attribute more events to the Poisson process.

Similarly, the estimates of $\alpha$, in conjunction with the estimates of $\beta$, for the backward process are overestimated. This  also suggests that for the backward process too much importance is given to the short term effect or impact of the previous events, and that the memory extends less into the history of the process.
It is worth noting that since the estimations of $\alpha$ and $\beta$ are similarly overestimated, the resulting estimates of the endogeneity $n$ is relatively close to the true value. Finally, closer to criticality there is an apparent tendency of the estimates for both arrows of time to converge.

It is also worth mentioning here that if we compare the forward and backward likelihood calculated with the MLE parameters for medium-size data sets (around $10^4$ events) we see that in 1.3\% of the cases that the backward likelihood actually is larger, and for even smaller data sets (around $500$ events) it is $16\%$. In practice, available data sets are typically quite small, and therefore the log-likelihood is not a guaranteed way to distinguish between the two arrows of time.

\subsection{Goodness of fit test}
For a given kernel $K$, baseline intensity $\lambda_0$ and a time series $\{t_i\}_{i=1,\dots,n}$ one defines the compensators 
\begin{equation}
\Lambda(t_{i-1},t_i)=\int\limits_{t_{i-1}}^{t_i}\lambda(s)\mathrm{d}s=\int\limits_{t_{i-1}}^{t_i}(\lambda_0+\sum\limits_{t_k<s}K(s-t_k))\mathrm{d}s
\label{eq:compensators}
\end{equation}
which are exponentially distributed with an average rate of 1 if the data comes from a HP \cite{papangelou1972integrability}. Thus we choose here the Kolmogorov-Smirnov test (KS test hereafter) to test the equality between the distribution of the compensators and the exponential distribution. The same test was used to find statistically valid fits of HP to high frequency data both in the foreign exchange market \cite{lallouache2016limits} and in the equity market \cite{omi2017hawkes}. 

\begin{figure}
\centering
\includegraphics[width=0.9\columnwidth]{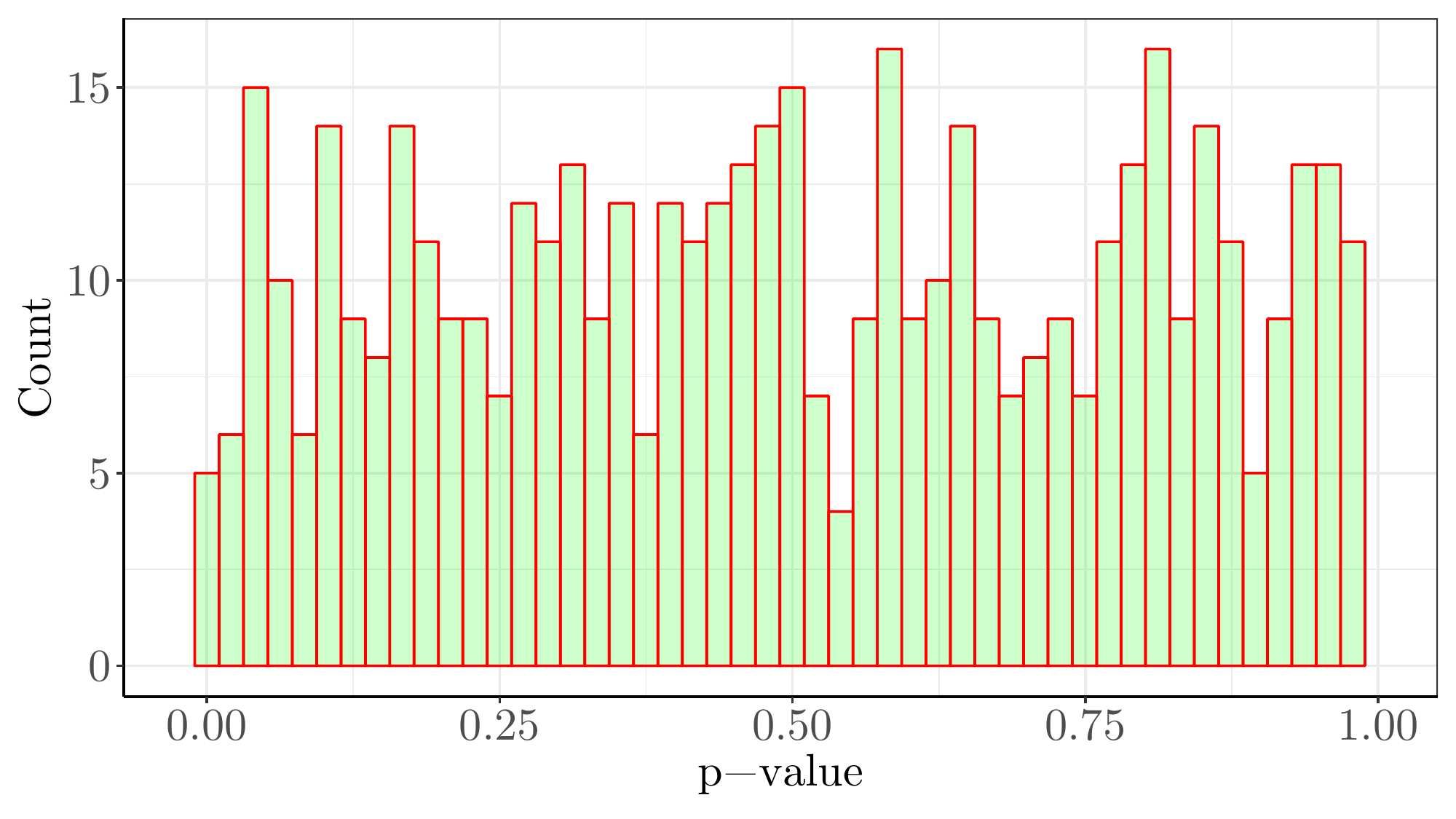}
\includegraphics[width=0.9\columnwidth]{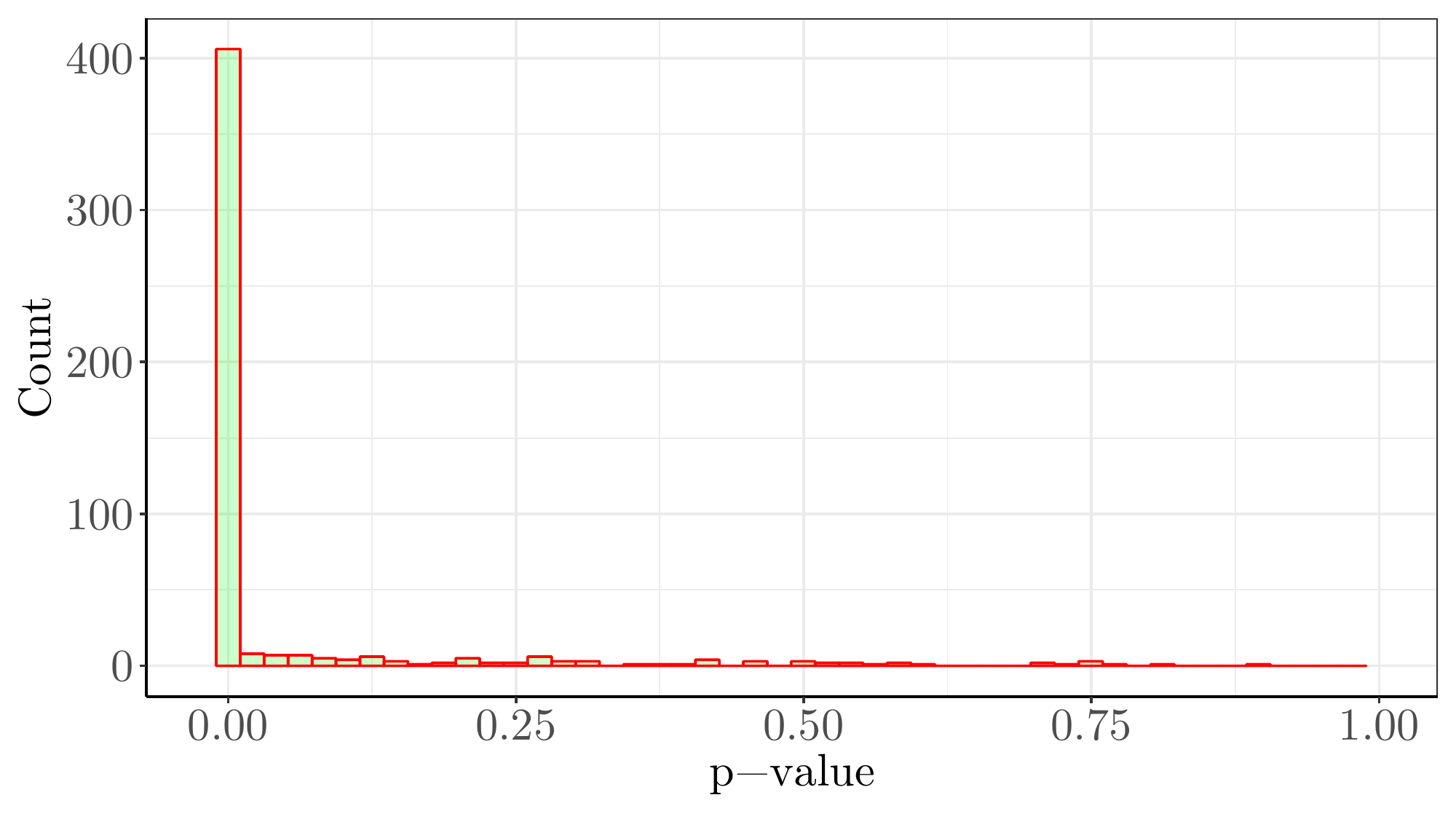}
\caption{Histogram of the p-values from the KS-test obtained for the forward (upper) and backward (lower) exponential HP with the true parameter values. The parameters used for the simulations are fixed to $\lambda_0=0.001$ and $\alpha=0.01$, with $\beta$ chosen to the desired endogeneity $n=\{0.50,0.75,0.90,0.95,0.99\}$. The data is collected over 100 runs for each parameter permutation and the expected number of events is set to $10^6$. }
\label{fig:histo_HP_exp_true_param}
\end{figure}

\begin{figure}
\centering
\includegraphics[width=0.9\columnwidth]{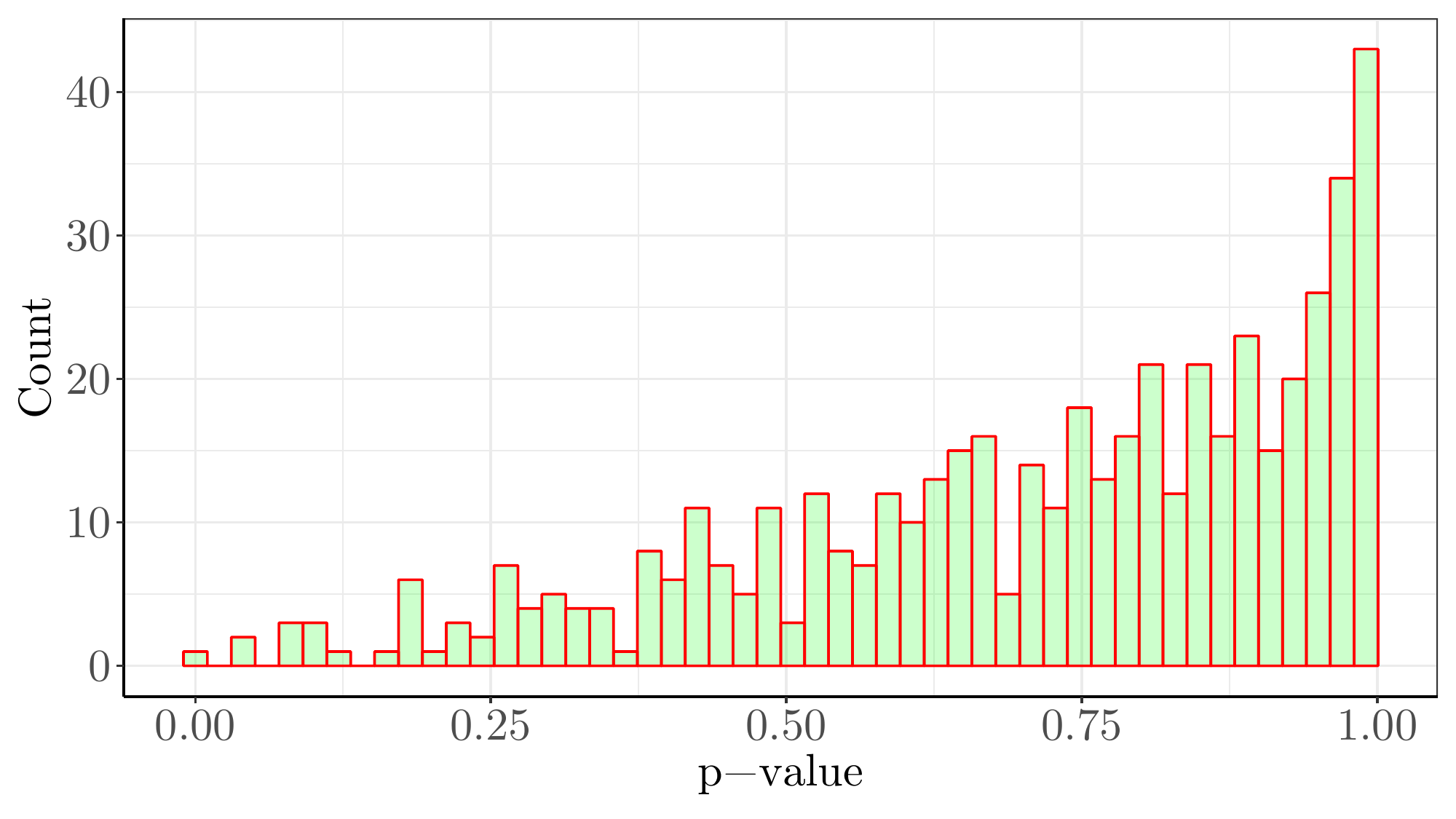}
\includegraphics[width=0.9\columnwidth]{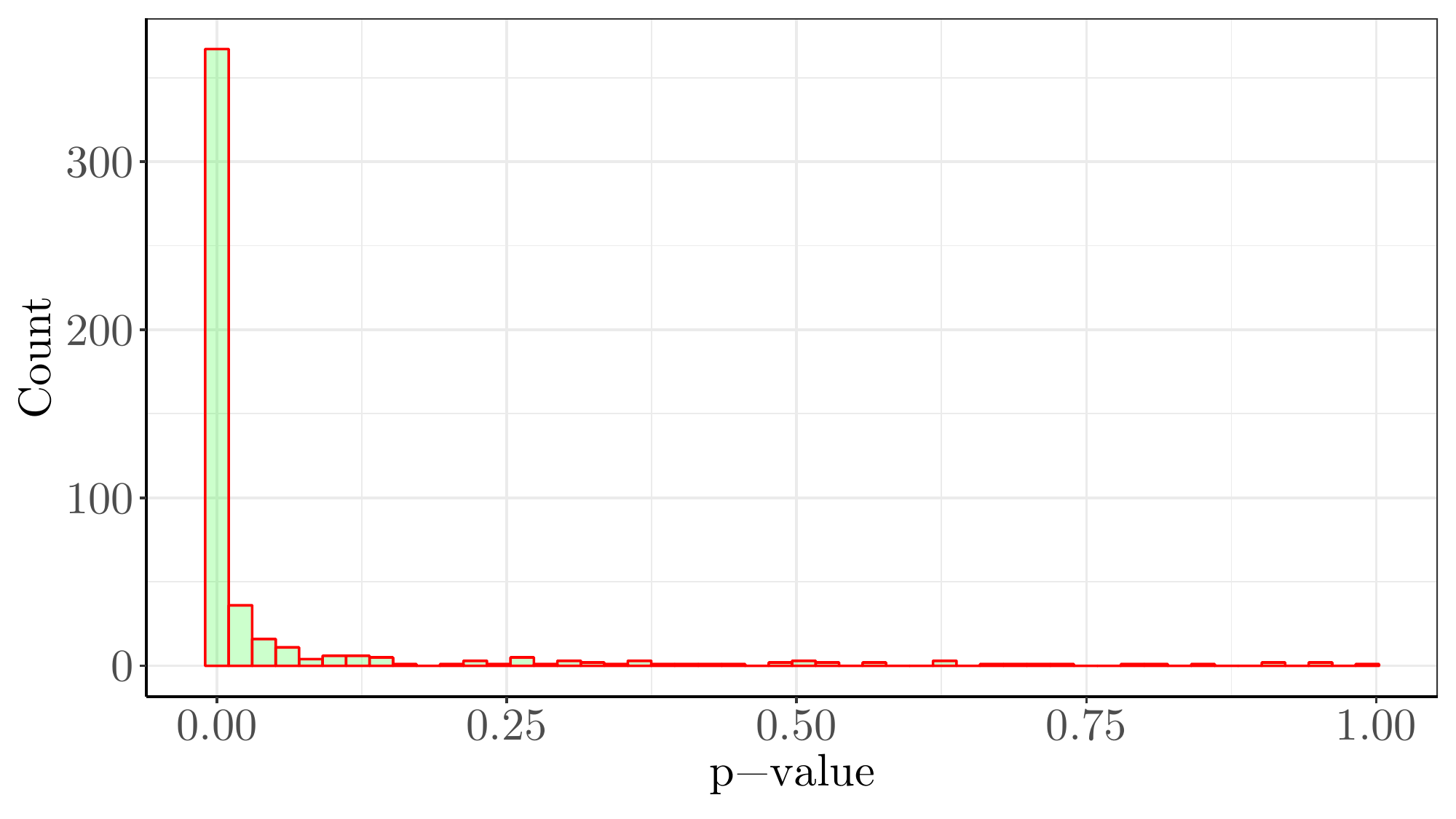}
\caption{Histogram of the p-values from the KS-test obtained for the forward (upper) and backward (lower) exponential HP with the MLE parameter values. The parameters are fixed to $\lambda_0=0.001$ and $\alpha=0.01$, with $\beta$ chosen to the desired endogeneity $n=\{0.50,0.75,0.90,0.95,0.99\}$. 100 runs for each parameter combination; expected number of events set to $10^6$. }
\label{fig:histo_HP_exp_MLE_param}
\end{figure}

Let us start with parametric estimation. We first test if the estimated kernel corresponds to the true one, i.e., the kernel obtained with the a priori known true parameter values ($\mu^*$, $\alpha^*$ and $\beta^*$). Figure \ref{fig:histo_HP_exp_true_param} displays the histogram of the p-values corresponding to this hypothesis. As expected for the forward case (upper plot), a uniform distribution is obtained since the null hypothesis holds. In the backward case, most fits are rejected. In a real-life situation, however, one does not know the true kernel. In this case, as shown by Fig.~\ref{fig:histo_HP_exp_MLE_param} where the parameters obtained via MLE are used ($\hat{\mu}_\textrm{ML}$, $\hat{\alpha}_\textrm{ML}$ and $\hat{\beta}_\textrm{ML}$), the test accepts more samples as being HPs processes, for both arrows of time. This is due to the additional freedom one has to find slightly over-fitting parameters. 

Thus, we see that the KS-test performs satisfactorily in the sense that it is clearly able to distinguish between the forward and backward process both for the MLE parameters (where in a sense the MLE "overfits" the parameters to the underlying data) and the true parameters. This emphasizes the need to assess the goodness of fits when fitting HPs to data. 

\begin{figure}
\centering
\includegraphics[width=0.9\columnwidth]{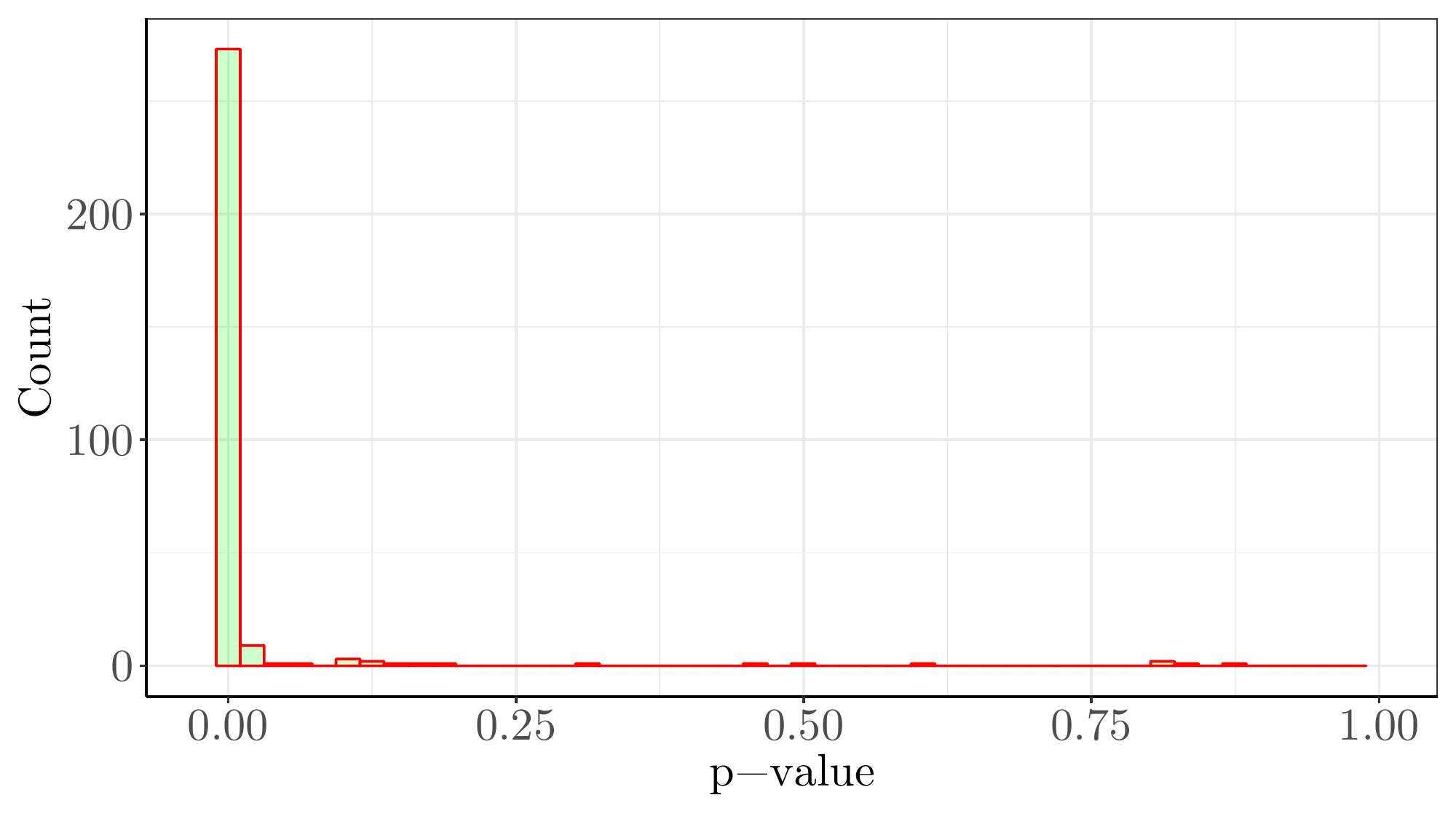}
\includegraphics[width=0.9\columnwidth]{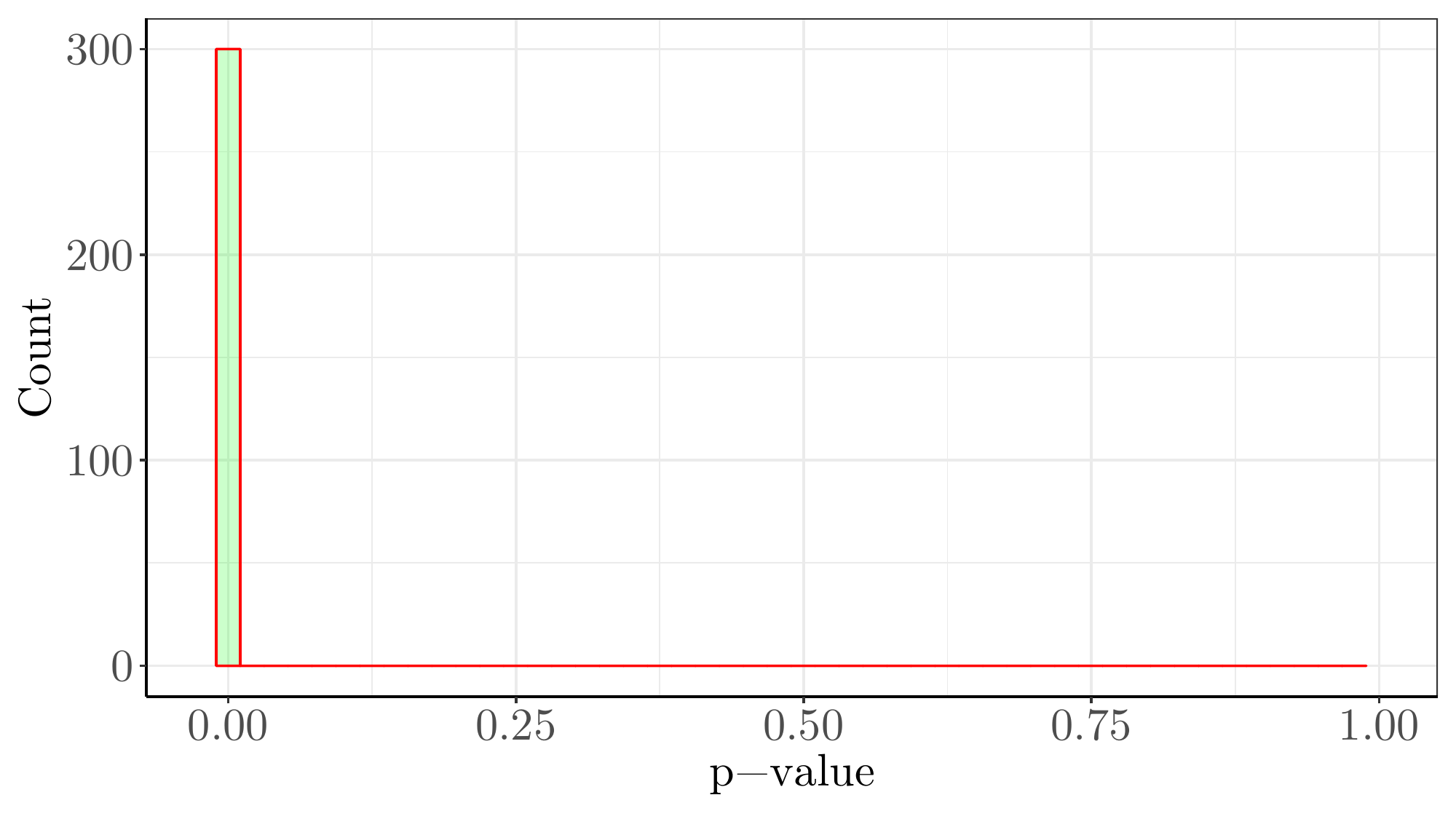}
\caption{Histogram of the p-values from the KS-test obtained for the forward (upper) and backward (lower) exponential HP with parameters extracted from the non-parametrically estimated kernels. The parameters are fixed to $\lambda_0=0.001$ and $\alpha=0.01$, with $\beta$ chosen to the desired endogeneity $n=\{0.50,0.75,0.90,0.95,0.99\}$. 100 runs for each parameter combination; expected number of events set to $10^6$.}
\label{fig:histo_HP_exp_non_param_est_param}
\end{figure}

The fact that the KS test is able to discriminate between the two arrows of time has a clear implication for the non-parametric kernel estimation method introduced by 
\cite{bacry2012non}: since it is based on the auto-covariance of the event rate, which is time-symmetric by definition, this methods yields the same kernel for both directions of time. As a consequence, in view of the power of the KS test in that respect, it is understandable that this method does not yield kernels that may be deemed statistically significant, as shown by Fig.\ \ref{fig:histo_HP_exp_non_param_est_param}, where the parameters used ($\hat{\mu}_\textrm{NP}$, $\hat{\alpha}_\textrm{NP}$ and $\hat{\beta}_\textrm{NP}$) are estimated from the non-parametrically obtained kernels by linear interpolation. More specifically, by taking the logarithm of the non-negative kernel estimate, estimates of $\hat{\alpha}_\textrm{NP}$ and $\hat{\beta}_\textrm{NP}$ are obtained by linear regression, and by Eq. \ref{eq:exp_no_events} we may obtain an estimate of $\hat{\mu}_\textrm{NP}$. It is worth noting here that this method, which is quite crude, produces a considerable amount of invalid results for the higher endogeneities (hence the smaller sample in Fig.~\ref{fig:histo_HP_exp_non_param_est_param}). 

We stress nevertheless that the non-parametric method is an invaluable tool to assess the global shape of HPs in a preliminary exploration, and to choose a suitable parametric family which itself may pass goodness of fit tests.

\subsection{Multivariate processes}

The above findings generalize to multivariate HPs, in which several univariate HPs may also mutually excite each other. More precisely, an $M$-dimensional ($M$ components) HP is defined as
\begin{equation}
\boldsymbol{\lambda}(t)=\boldsymbol{\lambda_0}+\int\limits_0^t\boldsymbol{G}(t-s)\mathrm{d}\boldsymbol{N_s}
\end{equation}
where the (exponential) kernel is given by 
\begin{equation}
G(t)=\left(\alpha^{mn}\mathrm{e}^{-\beta^{mn}(t-s)}\right)_{m,n=1,\dots,M}. 
\end{equation}

The intensity may thus be written as (with a constant baseline intensity) 
\begin{equation}
\lambda^m(t)=\lambda_0^m+\sum\limits_{n=1}^M\sum\limits_{t_i^n<t}\alpha^{mn}\mathrm{e}^{-\beta^{mn}(t-t_i^n)}.
\end{equation}
The expected number of events is
\begin{equation}
E\left[\boldsymbol{N}(t)\right]=\boldsymbol{\mu}t,
\end{equation}
where
\begin{equation}
\boldsymbol{\mu}=\left(\boldsymbol{I}-\int\limits_0^\infty \boldsymbol{G} (u)\mathrm{d}u\right)^{-1}\boldsymbol{\lambda_0}.
\end{equation} 
Here we define $N_t=\sum\limits_{m=1}^MN_t^m.$

For the multidimensional HP, denoting $\{t_i\}_{i=1,\dots,N}$ the ordered pool of all events $\{\{t_i^m\}_{m=1,\dots,M}\}$, the log-likelihood can be computed as the sum of the likelihood of each coordinate, namely
\begin{equation}
\ln\mathcal{L}\left(\{t_i\}_{i=1,\dots,N}\right)=\sum\limits_{m=1}^M\ln\mathcal{L}^m\left(\{t_i\}\right),
\end{equation}
where
\begin{equation}
\ln\mathcal{L}^m\left(\{t_i\}\right)=-\int\limits_0^T\lambda^m(s)\mathrm{d}s+\int\limits_0^T\ln\lambda^m(s)\mathrm{d}N^m_s.
\label{eq:log_lik_m_mHP}
\end{equation} 
Equation~(\ref{eq:log_lik_m_mHP}) may be written as 
\begin{equation}
\begin{aligned}
\ln\mathcal{L}^m\left(\{t_i\}\right)=-\lambda_0^mT
-\sum\limits_{n=1}^M\sum\limits_{t^n_i}\frac{\alpha^{mn}}{\beta^{mn}}\left(1-\mathrm{e}^{-\beta^{mn}(T-t^n_i)}\right)\\
+\sum\limits_{t_i^m}\ln\left[\lambda_0^m+\sum\limits_{n=1}^M\sum\limits_{t_k^n<t_i^m}\alpha^{mn}\mathrm{e}^{-\beta^{mn}(t_i-t_k^n)}\right].
\end{aligned}
\end{equation}

If we, as in the one-dimensional case, remove the non-stationary part of the process, we obtain
\begin{equation}
\begin{aligned}
\ln\mathcal{L}^m\left(\{t_i\}\right)=-\lambda_0^mT+\frac{\mu^m-\lambda_0^m}{\sum\limits_{n=1}^M\alpha^{mn}}\sum\limits_{n=1}^{M}\frac{\alpha^{mn}}{\beta^{mn}}\left(\mathrm{e}^{-\beta^{mn} T}-1\right)\\
-\sum\limits_{n=1}^M\sum\limits_{t^n_i}\frac{\alpha^{mn}}{\beta^{mn}}\left(1-\mathrm{e}^{-\beta^{mn}(T-t_i)}\right)\\
+\sum\limits_{t_i^m}\ln\left[\lambda_0^m+\left(\mu^m-\lambda_0^m\right)\frac{\sum\limits_{n=1}^M\alpha^{mn}\mathrm{e}^{-\beta^{mn}t_i}}{\sum\limits_{n=1}^M\alpha^{mn}}\right.\\
\left.+\sum\limits_{n=1}^M\sum\limits_{t_k^n<t_i}\alpha^{mn}\mathrm{e}^{-\beta^{mn}(t_i-t_k^n)}\right],
\end{aligned}
\end{equation}
 
Analogously to the univariate case, we can find an appropriate limit to the non-stationary period by considering
\begin{equation}
t_b=\inf\left\{t\in\{t_i\}_{i=1,\dots,N}:\sum\limits_{m=1}^M\lambda^m(t)\geq\sum\limits_{m=1}^M\mu^m\right\}.
\end{equation}
The process is then shifted $t_i^{m'}=t_i^m-t_b, t_i^m>t_b$ and $T'=T-t_b$.

A sufficient condition for stationarity is
\begin{equation}
\rho(\Gamma)=\max_{a\in S(\Gamma)} |a|<1,
\end{equation}
where $S(\Gamma)$ denotes the set of all eigenvalues of $\Gamma$ and
\begin{equation}
\Gamma=\int\limits_0^\infty\boldsymbol{G}(u)\mathrm{d}u=\left(\frac{\alpha^{mn}}{\beta^{mn}}\right)_{m,n=1\dots,M}
\end{equation}

Here we focus on symmetric multivariate HPs, where the mutual excitation matrix can be written as
\begin{displaymath} \boldsymbol{\alpha} = \left( \begin{array}{ccc} \alpha_0 & \alpha_m \\
\alpha_m & \alpha_0 \\
\end{array} \right), \end{displaymath}
For the sake of simplicity, we fix the baseline intensities and timescales to the same values for both components of the process, i.e., $\boldsymbol\lambda_0=(\lambda_0,\lambda_0)$ and $\boldsymbol{\beta}=  (\beta,\beta)$.

In the symmetric case $\rho(\Gamma)=\frac{\alpha_0+\alpha_m}{\beta}$. 
For the presentation of the results in the multivariate setting, the largest eigenvalue $\rho(\Gamma)$ was chosen as the control parameter instead of the endogeneity $n$ since it is directly linked to the expected total number of events in the process (i.e., summed over all components).

\begin{figure}
\centering
\includegraphics[width=0.9\columnwidth]{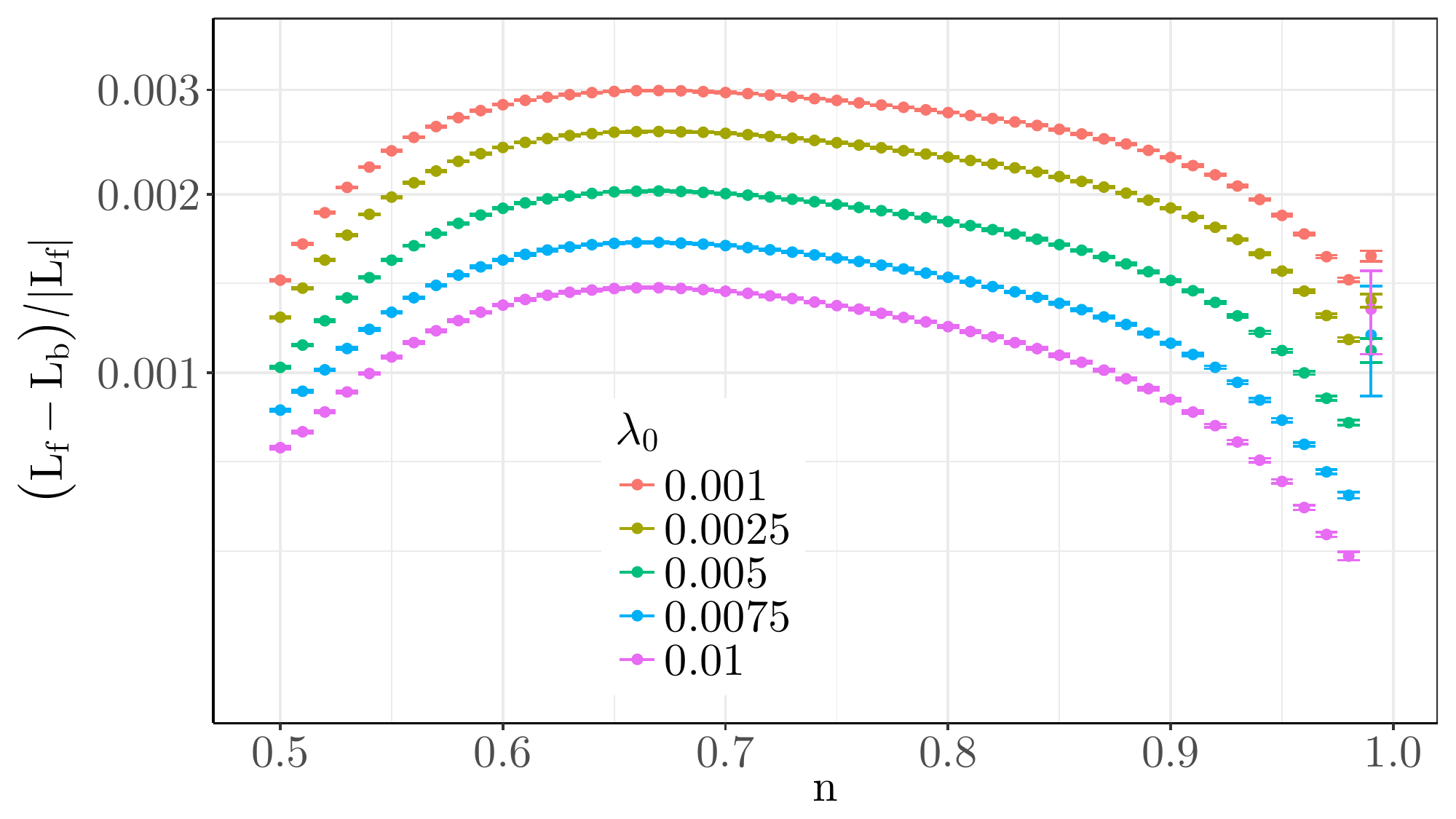}
\includegraphics[width=0.9\columnwidth]{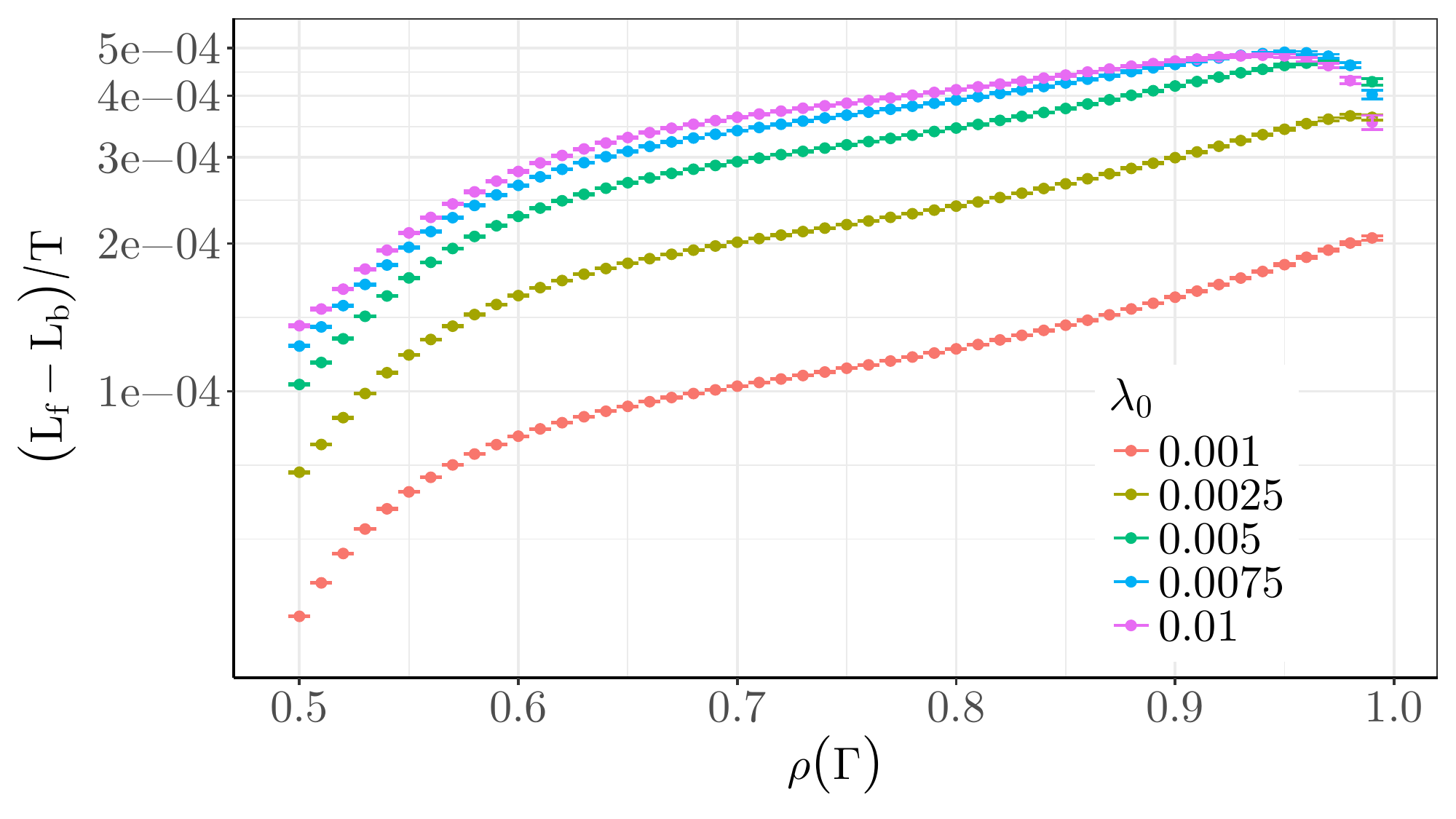}
\caption{Relative difference of the log-likelihood between forward and backward time arrows (top) and difference of the log-likelihood between forward and backward time arrows with regards to $T$ (bottom) for a multidimensional HP with a symmetric excitation kernel. All possible permutations of $\lambda_0=\{0.0010, 0.0025, 0.005, 0.0075, 0.0100\}$, $\alpha_0=\{0.049\}$, with $\alpha_m$ chosen according to the desired maximum eigenvalue $\rho(\Gamma)$, and $\beta=0.1$ are considered. The data points are grouped according to maximum eigenvalue and averaged over 100 runs for each parameter permutation. The expected total number of events is set to $10^6$.}
\label{fig:loglik_reldiff_mHP_symm}
\end{figure}

Figures~\ref{fig:loglik_reldiff_mHP_symm}, \ref{fig:fig_par_mHP_symm_full_alpha_m}, and \ref{fig:fig_par_mHP_symm_full_alpha_m_cont} show that the results for symmetric multidimensional HP are in many ways analogous to those of the univariate HP, i.e., the log-likelihood plots display a similar behaviour and the parametric estimations do not deviate significantly from each other in the forward and backward case.  
 Our remarks regarding the non-parametric method of \cite{bacry2012non}, which is only valid for symmetric HPs, still hold.

The above findings are however not true for asymmetric multivariate HPs, in which changing the direction of time leads to clearly different log-likelihoods and parameters (see Appendix \ref{appendix:asymmetric}), hence significantly increases  the effective causality of such processes.

\begin{figure}
\includegraphics[width=0.9\columnwidth]{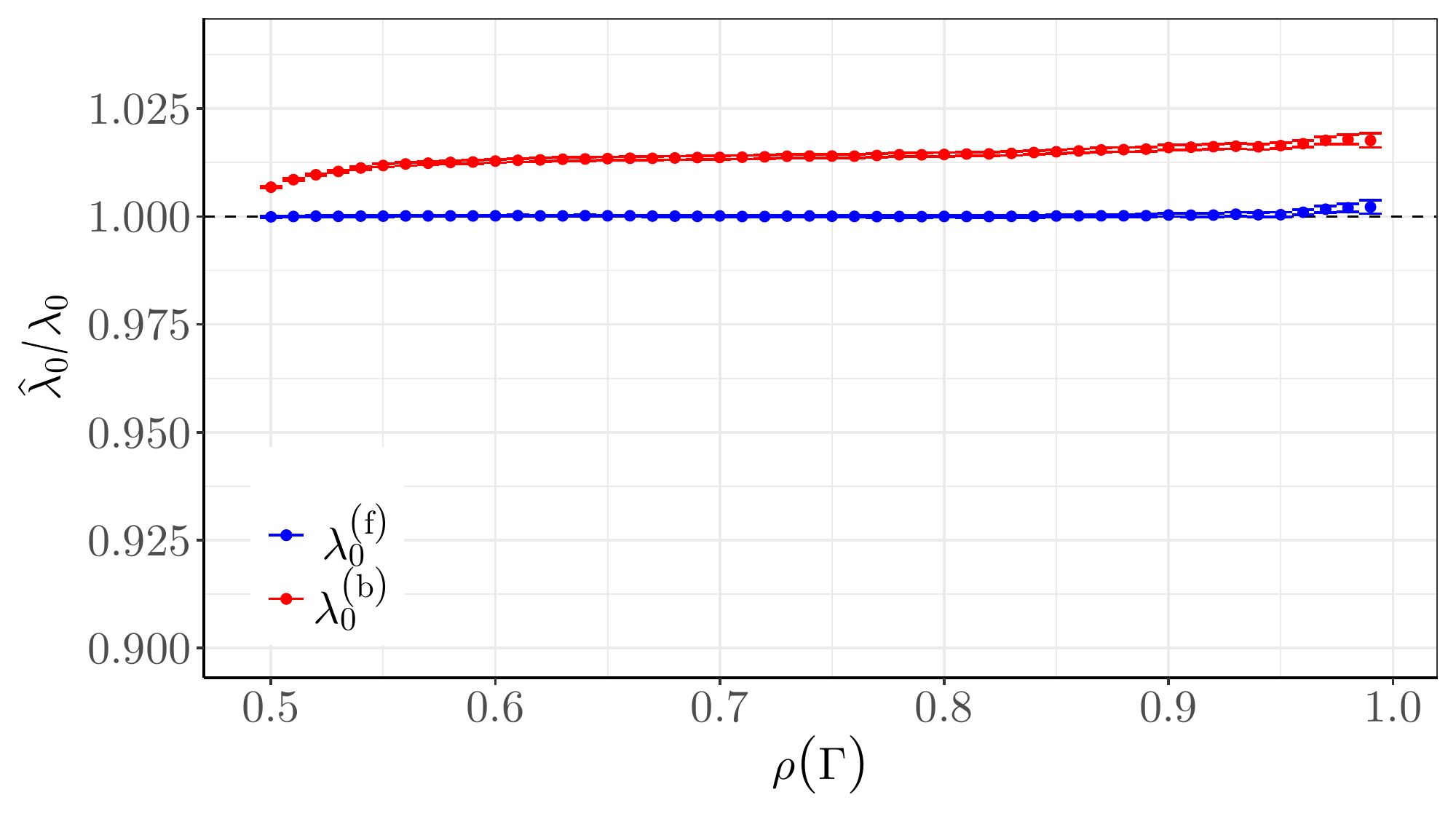}
\includegraphics[width=0.9\columnwidth]{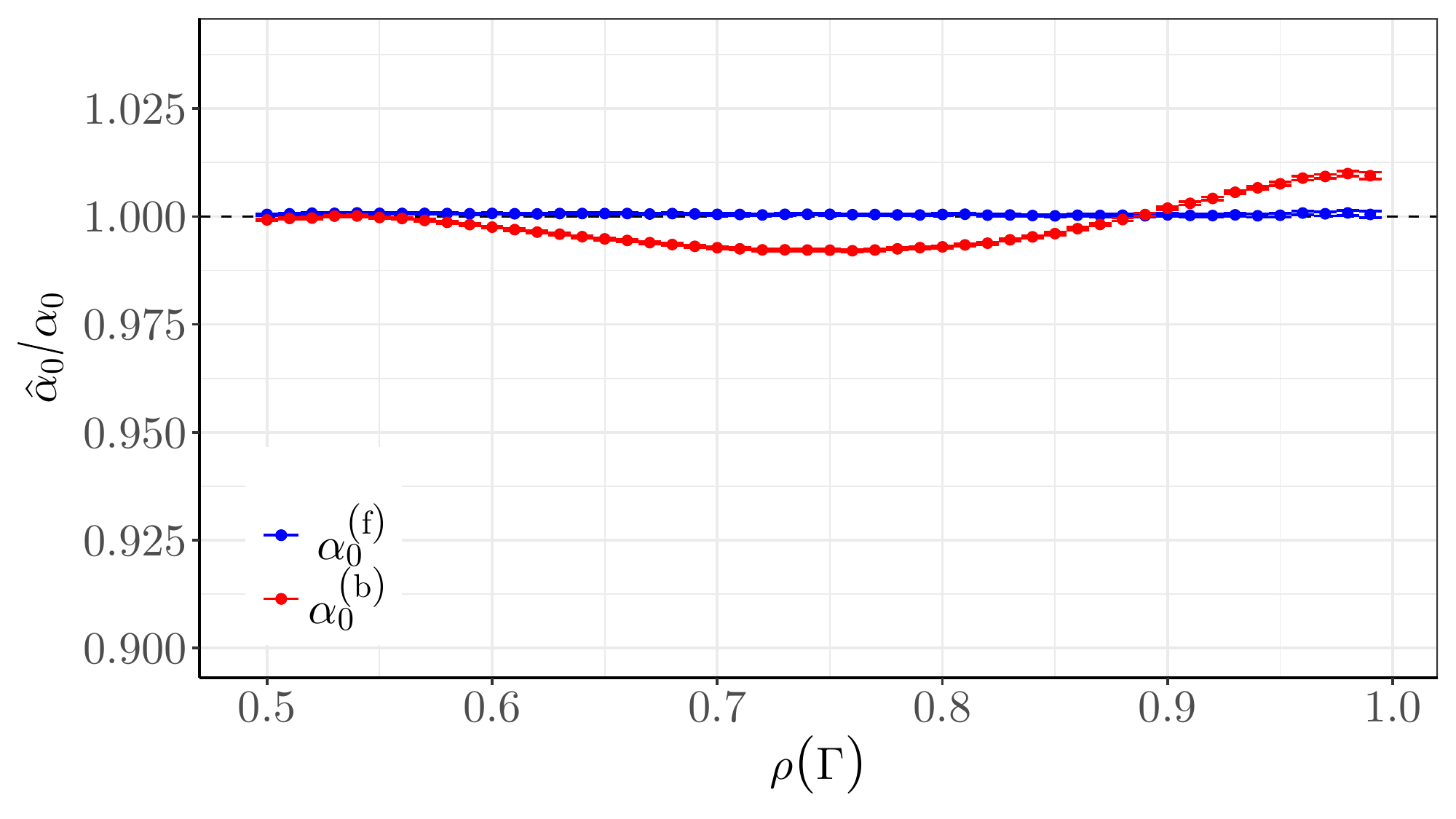}
\includegraphics[width=0.9\columnwidth]{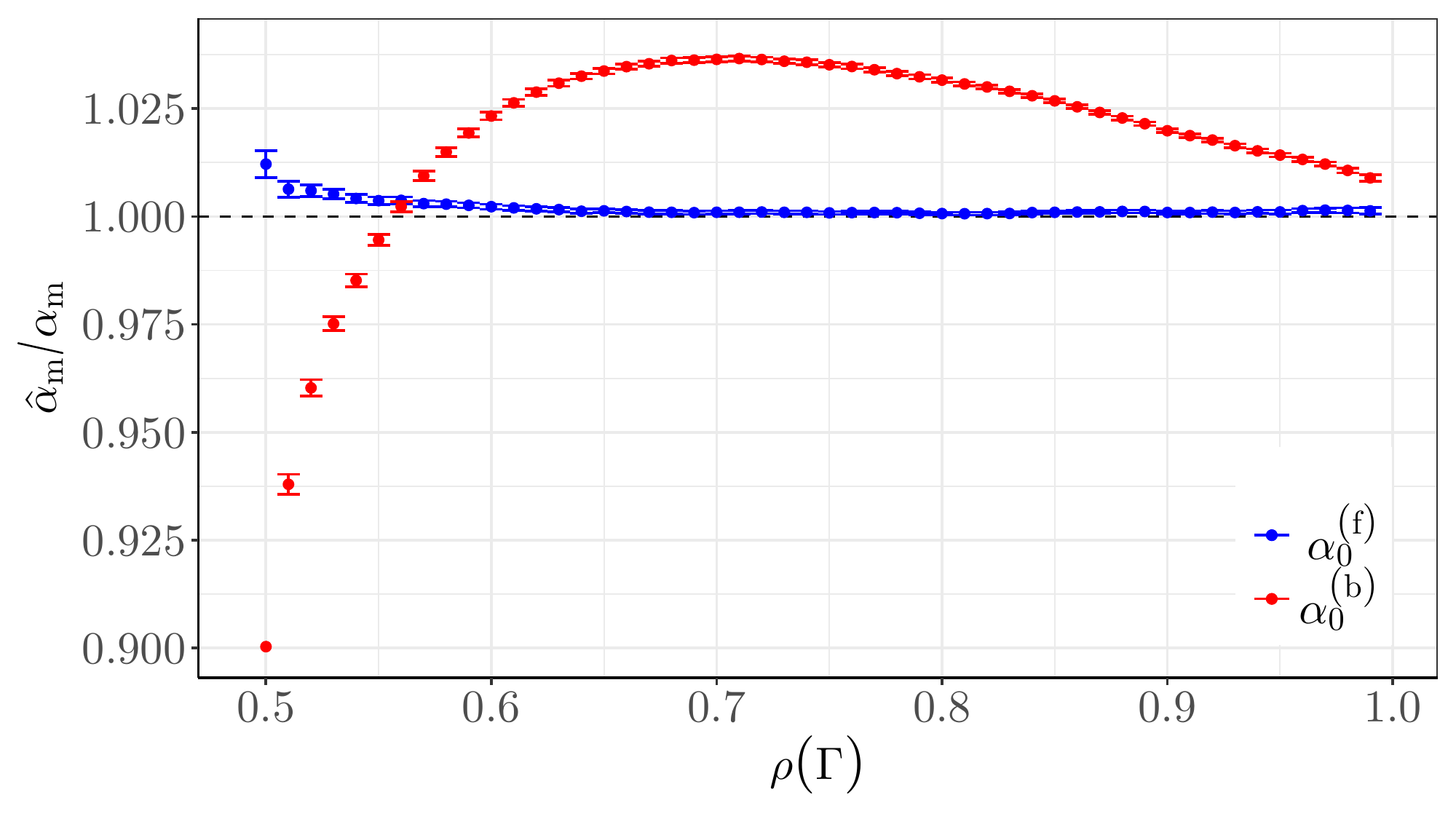}
\includegraphics[width=0.9\columnwidth]{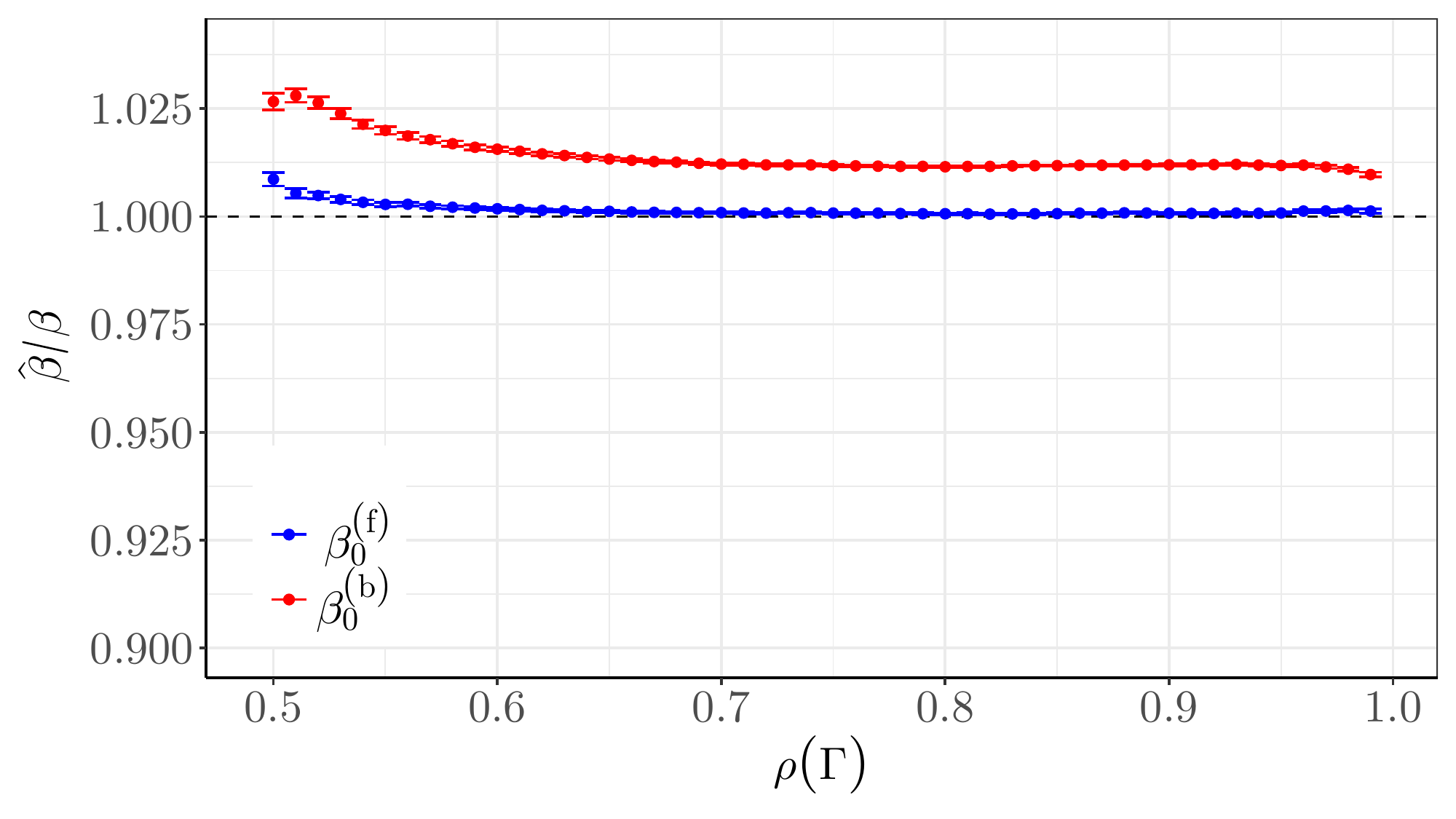}
\caption{Relative difference in the estimation of the various parameters in the MLE of the multidimensional HP with a symmetric excitation matrix for the forward (blue) and the backward process (red). All possible permutations of $\lambda_0=\{0.0010, 0.0025, 0.005, 0.0075, 0.0100\}$, $\alpha_0=\{0.049\}$, with $\alpha_m$ chosen according to the desired maximum eigenvalue $\rho(\Gamma)$, and $\beta=0.1$ are considered. The data points are grouped according to maximum eigenvalue and averaged over 100 runs for each parameter permutation. The expected total number of events is set to $10^6$.}
\label{fig:fig_par_mHP_symm_full_alpha_m}
\end{figure}

\begin{figure}
\centering
\includegraphics[width=0.9\columnwidth]{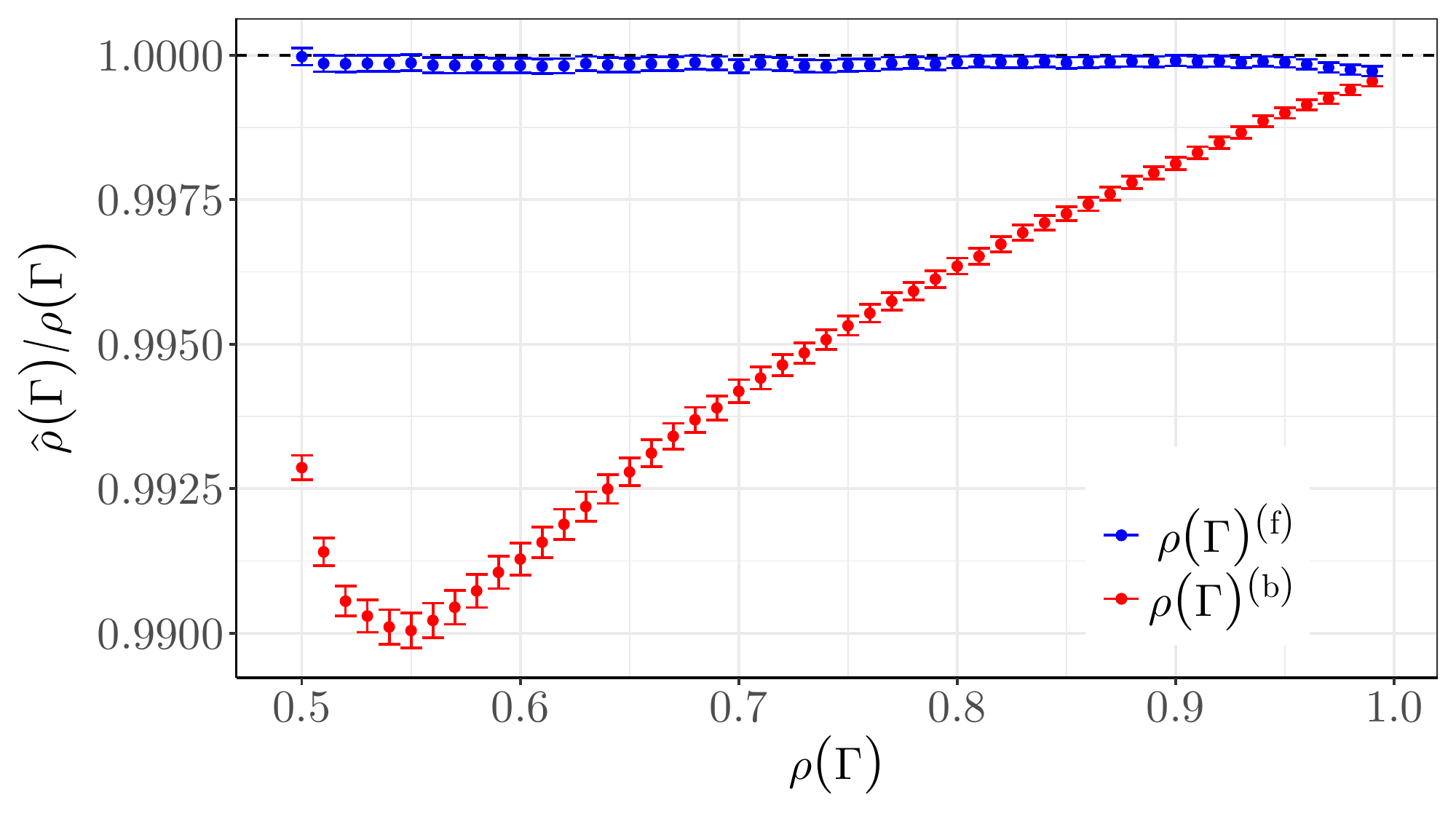}
\caption{(Continued) Relative difference in the estimation of the various parameters in the MLE of the multidimensional HP with a symmetric excitation matrix for the forward (blue) and the backward process (red). All possible permutations of $\lambda_0=\{0.0010, 0.0025, 0.005, 0.0075, 0.0100\}$, $\alpha_0=\{0.049\}$, with $\alpha_m$ chosen according to the desired maximum eigenvalue $\rho(\Gamma)$, and $\beta=0.1$ are considered. The data points are grouped according to maximum eigenvalue and averaged over 100 runs for each parameter permutation. The expected total number of events is set to $10^6$.}
\label{fig:fig_par_mHP_symm_full_alpha_m_cont}
\end{figure}

\section{Application to Data}
\label{sec:application_to_data}
Since the difference between forward and backward estimates is related to the endogeneity of the process, it is worth discussing some typical values found empirically. As an example of an application of the HP, we studied some fits of the HP to price data of the Exchange-Traded Fund SPDR S\&P 500 ETF. We follow largely the methods developed in \cite{lallouache2016limits}, and we focus on shorter time intervals where the authors find that the HP excel at fitting (one hour or less), i.e., time-intervals where we may assume that the baseline intensity is constant.

The data set encompasses price data over one day, \mbox{15-12-2015} from 9:30 to 16:00, and consists of approximately $950\,000$ data points. We tried the same fitting procedure for 23 other days and the results are consistent. We focus on bid prices and only consider changes in the bid price, which effectively reduces the dataset to only $33\,000$ data points. The coarse nature of the time has the interesting consequence that several bid price changes may occur during the same millisecond (the temporal resolution of the data). In order to address this issue, we assume that the order in which the price changes is correct and draw at random the times of the event within the same millisecond with a uniform distribution. One thus expects to lose some  causality because of the coarse temporal resolution.  

The fits are done with constant baseline intensity $\lambda_0$ and a kernel defined as a sum of exponentials, i.e., $K(t)=\sum\limits_{j=1}^{P}\alpha_j\mathrm{e}^{-\beta_jt}$, where $P=1,2,3$. This kind of kernel offers a lot of flexibility in terms of fitting. $P=2$ has been found to be a suitable choice for shorter time scales. We compare the results for both the non-stationary and stationary assumption (see Appendix~\ref{appendix:HP_sum_exp} for details), with, of course, a particular emphasis on what happens when the process is reversed. Unlike the case with the synthetic data, the time horizon is not known in the empirical data. It is thus assumed that the last event in the calibrated data set is the time horizon.  

The goodness of fits is not only assessed with KS test, but also with the Ljung-Box test (LB test), which checks if the compensators of HPs introduced in Eq.~(\ref{eq:compensators}) are not auto-correlated. We use the slight modification of the test introduced in~\cite{lallouache2016limits} in order to take into account the data cleaning procedure. The Aikaike Information Criterion (AIC) is also used to compare the merit of different kernels. The results for $P=1,2,3$ are presented in Table~\ref{table:trades_Nexp123}.

\begin{table}
\caption{Comparison of the ability of the exponential HP with $P=1,2,3$ to fit the empirical data (forward and backward) with different time windows. $n$ is the estimated endogeneity, $pKS$ and $pLB$ are the Kolmogorov-Smirnov and Ljung-Box test $p$-values, $\log(\mathcal{L})$ is the log-likelihood, $AIC$ is the Akaike Information Criterion and $N$ is the average number of events in a time window. Results obtained where the process is assumed to be stationary from the start are put in parenthesis. Values are averaged over all non-overlapping windows with more than 150 events.} 
\label{table:trades_Nexp123}
\small
\centering
\subcaption*{$P=1$ (forward)}
\scalebox{0.65}{
\begin{tabular}{|c|c|c|c|c|c|c|}
        \hline
         &  $n^{(f)}$ & $pKS^{(f)}$ & $pLB^{(f)}$ & $\log\mathcal{L}^{(f)}$ & $AIC^{(f}$ & $N$\\
        \hline
        1h & 0.33 (0.33) & 6.2e-6 (6.2e-6) & 8.3e-7 (8.3e-7) & 2282.04 (2282.08) & -4558.08 (-4558.15) & 4675\\
        \hline
        30m & 0.32 (0.32) & 1.3e-3 (1.3e-3) & 0.03 (0.03) & 1237.49 (1237.52) & -2468.99 (-2469.03) & 2518\\
        \hline
        15m & 0.32 (0.32) & 0.02 (0.02) & 0.06 (0.06) & 628.29 (628.31) & -1250.59 (-1250.61) & 1259\\
        \hline
        10m & 0.32 (0.32) & 0.07 (0.07) & 0.15 (0.15) & 420.01 (420.02) & -834.02 (-834.04) & 839\\
        \hline
        5m & 0.32 (0.32) & 0.24 (0.24) & 0.30 (0.30) & 212.95 (212.95) & -419.90 (-419.91) & 420\\
        \hline 
        \end{tabular}
}
\bigskip
\subcaption*{$P=1$ (backward)}
\scalebox{0.65}{
\begin{tabular}{|c|c|c|c|c|c|c|}
        \hline
         & $n^{(b)}$ & $pKS^{(b)}$ & $pLB^{(b)}$ & $\log\mathcal{L}^{(b)}$ & $AIC^{(b)}$ & $N$\\
        \hline
        1h & 0.32 (0.32) & 2.6e-7 (2.6e-7) & 2.6e-6 (2.6e-6) & 2262.78 (2262.78) & -4519.55 (-4519.57) & 4675\\
        \hline
        30m & 0.32 (0.32) & 3.9e-4 (3.9e-4) & 0.03 (0.03) & 1227.17 (1227.17) & -2448.35 (-2448.35) & 2518\\
        \hline
        15m & 0.31 (0.31) & 0.01 (0.01) & 0.05 (0.05) & 623.29 (623.28) & -1240.57 (-1240.57) & 1259\\
        \hline
        10m & 0.31 (0.31) & 0.04 (0.04) & 0.16 (0.16) & 416.61 (416.60) & -827.22 (-827.21) & 839\\
        \hline
        5m & 0.31 (0.31) & 0.19 (0.19) & 0.30 (0.30) & 211.29 (211.29) & -416.59 (-416.58) & 420\\
        \hline 
\end{tabular}
}

\bigskip
\subcaption*{$P=2$ (forward)}
\scalebox{0.65}{
\begin{tabular}{|c|c|c|c|c|c|c|}
        \hline
         &  $n^{(f)}$ & $pKS^{(f)}$ & $pLB^{(f)}$ & $\log\mathcal{L}^{(f)}$ & $AIC^{(f}$ & $N$\\
        \hline
        1h & 0.59 (0.59) & 0.02 (0.02) & 0.17 (0.17) & 2438.63 (2438.69) & -4867.26 (-4867.38) & 4675\\
        \hline
        30m & 0.57 (0.57) & 0.17 (0.17) & 0.34 (0.34) & 1317.87 (1317.92) & -2625.75 (-2625.85) & 2518\\
        \hline
        15m & 0.53 (0.53) & 0.40 (0.40) & 0.35 (0.35) & 665.25 (665.28) & -1320.51 (-1320.56) &  1259\\
        \hline
        10m & 0.54 (0.54) & 0.47 (0.47) & 0.38 (0.37) & 444.61 (444.58) & -879.22 (-879.16) & 840\\
        \hline
        5m & 0.51 (0.51) & 0.64 (0.63) & 0.43 (0.44) & 225.42 (224.72) & -440.83(-439.43) & 420\\
        \hline
        \end{tabular}
        }
\bigskip
\small
\centering
\subcaption*{$P=2$ (backward)}
\scalebox{0.65}{
\begin{tabular}{|c|c|c|c|c|c|c|}
        \hline
         & $n^{(b)}$ & $pKS^{(b)}$ & $pLB^{(b)}$ & $\log\mathcal{L}^{(b)}$ & $AIC^{(b)}$ & $N$\\
        \hline
        1h & 0.59 (0.59) & 3.8e-4 (3.9e-4) & 0.18 (0.18) & 2386.57 (2386.59) & -4763.14 (-4763.18) & 4675\\
        \hline
        30m & 0.55 (0.55) & 0.03 (0.03) & 0.32 (0.32) & 1290.59 (1290.60) & -2571.18 (-2571.21) & 2518\\
        \hline
        15m & 0.49 (0.49) & 0.16 (0.17) & 0.31 (0.30) & 651.78 (652.14) & -1293.56 (-1294.27) & 1259\\
        \hline
        10m & 0.50 (0.50) & 0.26 (0.26) & 0.36 (0.36) & 435.90 (436.16) & -861.80 (-862.32) & 840\\
        \hline
        5m & 0.47 (0.47) & 0.48 (0.48) & 0.40 (0.41) & 220.40 (220.37) & -430.79 (-430.73) & 420\\
        \hline
        \end{tabular}
        }
\bigskip
\subcaption*{$P=3$ (forward)}
\scalebox{0.65}{
\begin{tabular}{|c|c|c|c|c|c|c|}
        \hline
         &  $n^{(f)}$ & $pKS^{(f)}$ & $pLB^{(f)}$ & $\log\mathcal{L}^{(f)}$ & $AIC^{(f}$ & $N$\\
        \hline
        1h & 0.69 (0.69) & 0.01 (0.03) & 0.22 (0.21) & 2468.46 (2480.59) & -4922.92 (-4947.19) & 4675 \\
        \hline
        30m & 0.67 (0.66) & 0.15 (0.15) & 0.38 (0.38) & 1333.28 (1339.70) & -2652.57 (-2665.41) & 2518\\
        \hline
        15m & 0.60 (0.59) & 0.40 (0.39) & 0.40 (0.40) & 674.12 (674.11) & -1334.25 (-1334.22) & 1259 \\
        \hline
        10m & 0.60 (0.61) & 0.50 (0.49) & 0.44 (0.44) & 449.38 (448.05) & -884.77 (-882.09) & 839\\
        \hline
        5m & 0.55 (0.55) & 0.72 (0.71) & 0.46 (0.46) & 227.80 (227.80) & -441.59 (-441.60) & 419 \\
        \hline
        \end{tabular}
        }
\bigskip
\subcaption*{$P=3$ (backward)}
\scalebox{0.65}{
\begin{tabular}{|c|c|c|c|c|c|c|}
        \hline
         & $n^{(b)}$ & $pKS^{(b)}$ & $pLB^{(b)}$ & $\log\mathcal{L}^{(b)}$ & $AIC^{(b)}$ & $N$\\
        \hline
        1h & 0.67 (0.67) & 4.4e-03 (0.01) & 0.29 (0.28) & 2412.88 (2412.51) & -4811.76 (-4811.03) & 4675 \\
        \hline
        30m & 0.64 (0.63) & 0.07 (0.05) & 0.40 (0.41) & 1303.68 (1302.34) & -2593.35 (-2590.68) & 2518\\
        \hline
        15m & 0.57 (0.57) & 0.20 (0.20) & 0.39 (0.38) & 658.42 (658.41) & -1302.85 (-1302.82) & 1259 \\
        \hline
        10m &  0.55 (0.55) & 0.30 (0.31) & 0.40 (0.42) & 438.79 (439.00) & -863.58 (-863.99) & 839\\
        \hline
        5m & 0.52 (0.52) & 0.54 (0.53) & 0.44 (0.44) & 222.10 (222.15) & -430.20 (-430.30) & 419 \\
        \hline
        \end{tabular}
        }

\end{table}

It is clear that assuming that the process is stationary, and using the slightly modified methods, does not significantly improve the fits, and does not merit much attention. However, if we compare the forward and backward cases, we see that unlike when the synthetic data was considered, it is is not as clear cut and it is not sufficient to consider just the $p$-value obtained in the KS test to determine the arrow of time. The forward case does indeed consistently perform better than the backward case but the values obtained for the backward case are still acceptable, and when the degrees of freedom in the model are increased, we generally get a better $p$-value.

If we turn our attention to the LB test the situation is similar, but here the difference between the two cases is even smaller. In fact, we sometimes see that the backward process occasionally, when $P=3$, performs better than the the forward process. On the other hand the log-likelihood is consistently larger for the forward case, but not significantly. Finally, the AIC favours a kernel with a larger number of degrees of freedom.

The whole picture makes sense in the light of the results on synthetic data: the estimated endogeneity $n$ depends on the kernel and time window chosen and varies between approximately 0.30 and 0.70, hence far from criticality, a region in which  the difference between forward and backward results is small, hence causality is weak.

\section{Discussion}

The above findings for both the univariate and symmetric multivariate cases have several consequences. First, their causality is much weaker than previously implicitly assumed, even with synthetic data whose kernel family is known. This in turn makes it sometimes difficult to distinguish between the forward and backward event time vectors and thus strongly emphasises the importance of using goodness of fit tests, even for synthetic data. Since the kernel is the only causal term in HPs, weak causality contributes to the difficulties in fitting HPs, especially when the baseline intensity varies with time. This is one of the reasons why we have accounted for the potential lack of initial non-stationary part when calibrating HPs in some cases with a modified log-likelihood function such as the one we propose, a point which has received little attention to our knowledge.

A practical consequence of this work is that fitting weakly causal HPs to real data rests on even shallower ground because of data imperfection, for two main reasons. First, data cannot be assumed to be perfect; for example the time resolution of the data may be coarse enough to allow several events to take place during the same data time and the event times may be affected by non-negligible noise, as it happens sometimes in financial tick-by-tick data. These two time-related problems further weaken the causality of HPs. Second, the shape of the kernel is a priori unknown.

This raises an important issue: simple kernels with very few degrees of freedom are seldom satisfactory, thus more complex models with more degrees of freedom are introduced until satisfactory results are achieved. The same is true for the backward arrow of time, and reassuringly, the results are often worse, but not systematically and certainly not in a manner as convincing as when one knows the kernel shape. In other words, the larger the number of degrees of freedom of a kernel, the more successful the fits, but at the cost of weakening the difference between forward and backward arrows of time because more degrees of freedom also allow a more precise fit of the backward event times vector. At all rates, our results suggest an additional test for HPs: one should reject the hypothesis that HPs describe the data if the forward event time vector leads to worse goodness of fit tests than the backward event time vector. Indeed, in such cases, it makes little sense to trust the causality that HPs introduce.




\bibliographystyle{plain}
\bibliography{my_references}

\begin{thebibliography}{10}

\bibitem{bacry2012non}
Emmanuel Bacry, Khalil Dayri, and Jean-Fran{\c{c}}ois Muzy.
\newblock Non-parametric kernel estimation for symmetric hawkes processes.
  application to high frequency financial data.
\newblock {\em The European Physical Journal B-Condensed Matter and Complex
  Systems}, 85(5):1--12, 2012.

\bibitem{byrd1995limited}
Richard~H Byrd, Peihuang Lu, Jorge Nocedal, and Ciyou Zhu.
\newblock A limited memory algorithm for bound constrained optimization.
\newblock {\em SIAM Journal on Scientific Computing}, 16(5):1190--1208, 1995.

\bibitem{dassios2017generalized}
Angelos Dassios and Hongbiao Zhao.
\newblock A generalized contagion process with an application to credit risk.
\newblock {\em International Journal of Theoretical and Applied Finance},
  20(01):1750003, 2017.

\bibitem{filimonov2012quantifying}
Vladimir Filimonov and Didier Sornette.
\newblock Quantifying reflexivity in financial markets: Toward a prediction of
  flash crashes.
\newblock {\em Physical Review E}, 85(5):056108, 2012.

\bibitem{gardner1974sequence}
JK~Gardner and Leon Knopoff.
\newblock Is the sequence of earthquakes in southern california, with
  aftershocks removed, poissonian?
\newblock {\em Bulletin of the Seismological Society of America},
  64(5):1363--1367, 1974.

\bibitem{kirchner2017estimation}
Matthias Kirchner.
\newblock An estimation procedure for the hawkes process.
\newblock {\em Quantitative Finance}, 17(4):571--595, 2017.

\bibitem{lallouache2016limits}
Mehdi Lallouache and Damien Challet.
\newblock The limits of statistical significance of hawkes processes fitted to
  financial data.
\newblock {\em Quantitative Finance}, 16(1):1--11, 2016.

\bibitem{london2010sensitivity}
Michael London, Arnd Roth, Lisa Beeren, Michael H{\"a}usser, and Peter~E
  Latham.
\newblock Sensitivity to perturbations in vivo implies high noise and suggests
  rate coding in cortex.
\newblock {\em Nature}, 466(7302):123, 2010.

\bibitem{marsan2008extending}
David Marsan and Olivier Lengline.
\newblock Extending earthquakes reach through cascading.
\newblock {\em Science}, 319(5866):1076--1079, 2008.

\bibitem{mohler2011self}
George~O Mohler, Martin~B Short, P~Jeffrey Brantingham, Frederic~Paik
  Schoenberg, and George~E Tita.
\newblock Self-exciting point process modeling of crime.
\newblock {\em Journal of the American Statistical Association},
  106(493):100--108, 2011.

\bibitem{nelder1965simplex}
John~A Nelder and Roger Mead.
\newblock A simplex method for function minimization.
\newblock {\em The computer journal}, 7(4):308--313, 1965.

\bibitem{ogata1981lewis}
Yosihiko Ogata.
\newblock On lewis' simulation method for point processes.
\newblock {\em IEEE Transactions on Information Theory}, 27(1):23--31, 1981.

\bibitem{ogata1988statistical}
Yosihiko Ogata.
\newblock Statistical models for earthquake occurrences and residual analysis
  for point processes.
\newblock {\em Journal of the American Statistical association}, 83(401):9--27,
  1988.

\bibitem{omi2017hawkes}
Takahiro Omi, Yoshito Hirata, and Kazuyuki Aihara.
\newblock Hawkes process model with a time-dependent background rate and its
  application to high-frequency financial data.
\newblock {\em to appear in Phys. Rev. E}, 2017.

\bibitem{papangelou1972integrability}
F~Papangelou.
\newblock Integrability of expected increments of point processes and a related
  random change of scale.
\newblock {\em Transactions of the American Mathematical Society},
  165:483--506, 1972.

\bibitem{pillow2008spatio}
Jonathan~W Pillow, Jonathon Shlens, Liam Paninski, Alexander Sher, Alan~M
  Litke, EJ~Chichilnisky, and Eero~P Simoncelli.
\newblock Spatio-temporal correlations and visual signalling in a complete
  neuronal population.
\newblock {\em Nature}, 454(7207):995, 2008.

\bibitem{roueff2016locally}
Fran{\c{c}}ois Roueff, Rainer Von~Sachs, and Laure Sansonnet.
\newblock Locally stationary hawkes processes.
\newblock {\em Stochastic Processes and their Applications}, 126(6):1710--1743,
  2016.

\bibitem{soros2003alchemy}
George Soros.
\newblock {\em The alchemy of finance}.
\newblock John Wiley \& Sons, 2003.

\bibitem{truccolo2005point}
Wilson Truccolo, Uri~T Eden, Matthew~R Fellows, John~P Donoghue, and Emery~N
  Brown.
\newblock A point process framework for relating neural spiking activity to
  spiking history, neural ensemble, and extrinsic covariate effects.
\newblock {\em Journal of neurophysiology}, 93(2):1074--1089, 2005.

\bibitem{zhuang2002stochastic}
Jiancang Zhuang, Yosihiko Ogata, and David Vere-Jones.
\newblock Stochastic declustering of space-time earthquake occurrences.
\newblock {\em Journal of the American Statistical Association},
  97(458):369--380, 2002.

\end{thebibliography}

\appendix
\section{Univariate Hawkes processes with power-law kernels}
\label{appendix:powerlaw}
\begin{figure}
\centering{
\includegraphics[width=0.9\columnwidth]{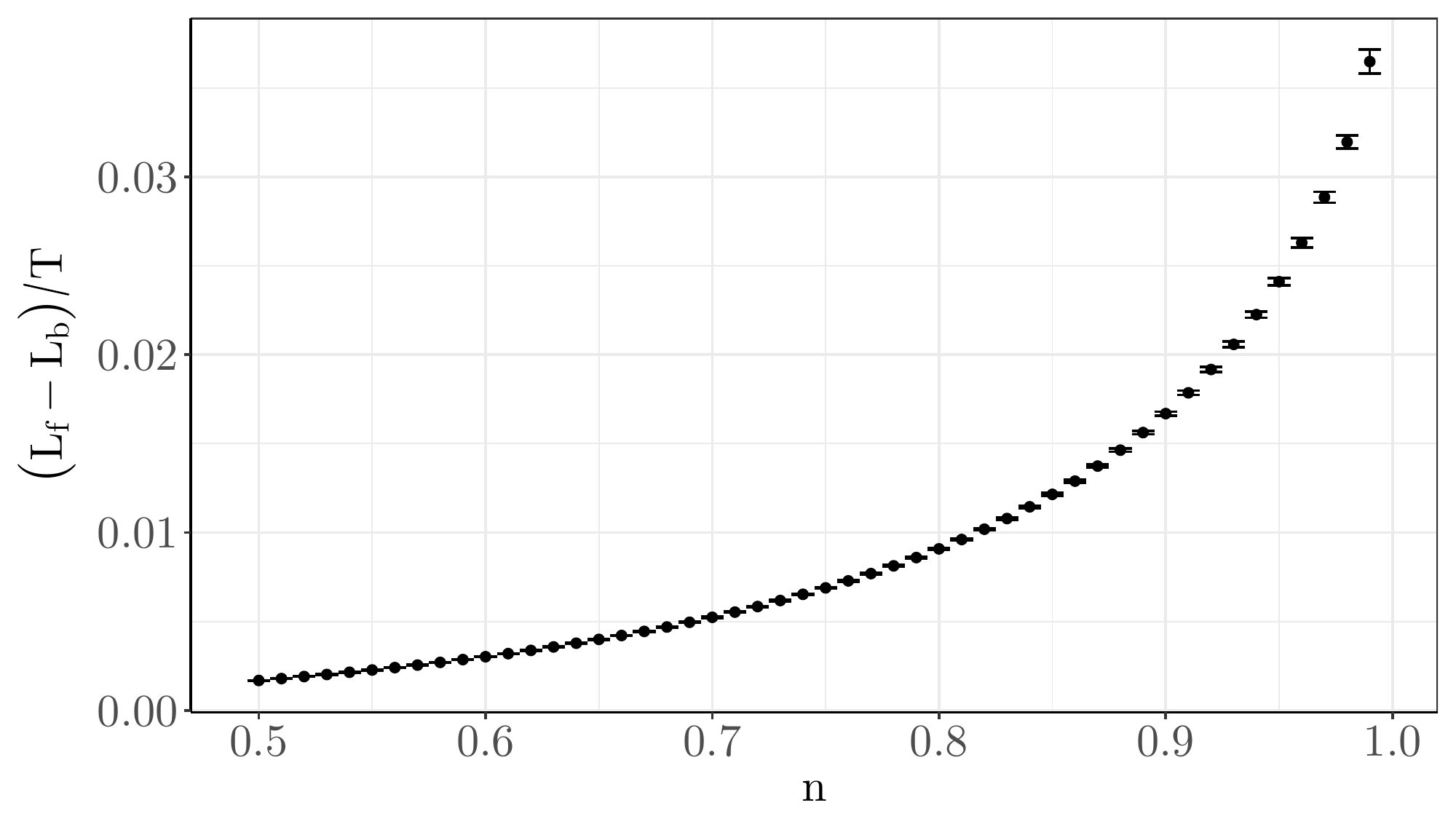}
}
\caption{Relative difference of the log-likelihood with regards to $T$ (the time horizon of the simulations) between forward and backward time arrows for a HP with a power-law kernel. The selection of parameters is limited to $\lambda_0=0.05$, $u=0.06$,$w=-2.5$ and with a varying $v$ chosen according to the desired endogeneity $n$. The data points are grouped according to their endogeneity and averaged over 100 runs. The expected number of events is set to $10^5$.}
\label{fig:loglik_reldiff_pow}
\end{figure}

A power-law kernel for the univariate HP may be defined as
\begin{equation}
K(t)=u(t+v)^{w}.
\label{eq:power_law_kernel}
\end{equation}
The endogeneity thus equals
\begin{equation}
n=-\frac{u}{w+1}v^{w+1}
\end{equation}
and the log-likelihood is given by
\begin{equation}
\begin{aligned}
\ln\mathcal{L}\left(\left\{t_i\right\}_{i=1,\dots,n }\right)=-\lambda_0T\\-\sum\limits_{i=1}^n\frac{u}{w+1}\left((T-t_i+v)^{w+1}-v^{w+1}\right)\\+\sum\limits_{i=1}^n\ln\left[\lambda_0+\sum\limits_{k=1}^{i-1}u(t_i-t_k+v)^{w}\right].
\end{aligned}.
\end{equation}
If the initial non-stationary part of the process is removed, the mathematical expression above must be modified;
\begin{equation}
\begin{aligned}
\ln\mathcal{L}\left(\left\{t_i\right\}_{i=1,\dots,n }\right)=-\lambda_0T\\
-\frac{v}{w+1}\left(\frac{\lambda_0}{1+\frac{u}{w+1}v^{w+1}}-\lambda_0\right)\left((\frac{T}{v}+1)^{w+1}-1\right)\\
-\sum\limits_{i=1}^n\frac{u}{w+1}\left((T-t_i+v)^{w+1}-v^{w+1}\right)\\+\sum\limits_{i=1}^n\ln\left[\lambda_0+\left(\frac{\lambda_0}{1+\frac{u}{w+1}v^{w+1}}-\lambda_0\right)\left(\frac{t_i}{v}+1\right)^w\right.\\
+\left.\sum\limits_{k=1}^{i-1}u(t_i-t_k+v)^{w}\right].
\end{aligned}
\end{equation}
Figure~\ref{fig:loglik_reldiff_pow} displays the relative difference of the log-likelihood as a function of the endogeneity. 

\section{Log-likelihood of the univariate HP with a sum of exponentials}
\label{appendix:HP_sum_exp}
If the HP kernel consists of a sum of $P$ exponentials, i.e.,
\begin{equation}
K(t)=\sum\limits_{j=1}^{P}\alpha_j\mathrm{e}^{-\beta_jt},
\end{equation}
then the associated log-likelihood is given by
\begin{equation}
\begin{aligned}
\ln\mathcal{L}\left(\left\{t_i\right\}_{i=1,\dots,n }\right)=-\lambda_0T-\sum\limits_{i=1}^n\sum\limits_{j=1}^{P}\frac{\alpha_j}{\beta_j}\left(1-\mathrm{e}^{\beta_j(T-t_i)}\right)\\
+\sum\limits_{i=1}^n\ln\left[\lambda_0+\sum\limits_{k=1}^{i-1}\sum\limits_{j=1}^{P}\alpha_j\mathrm{e}^{-\beta_j(t_i-t_k)}\right].
\end{aligned}
\end{equation}
Consequently, the modified log-likelihood, where it is assumed that there is no initial non-stationary part, is given by \begin{equation}
\begin{aligned}
\ln\mathcal{L}\left(\left\{t_i\right\}_{i=1,\dots,n }\right)=-\lambda_0T+\frac{\left(\frac{\lambda_0}{1-n}-\lambda_0\right)}{\sum\limits_{j=1}^{P}\alpha_j}\left(\sum\limits_{j=1}^P\frac{\alpha_j}{\beta_j}\left(\mathrm{e}^{-\beta_jT}-1\right)\right)\\
-\sum\limits_{i=1}^n\sum\limits_{j=1}^{P=2}\frac{\alpha_j}{\beta_j}\left(1-\mathrm{e}^{\beta_j(T-t_i)}\right)\\
+\sum\limits_{i=1}^n\ln\left[\lambda_0+\left(\frac{\lambda_0}{1-n}-\lambda_0\right)\left(\frac{\sum\limits_{j=1}^P\alpha_j\mathrm{e}^{-\beta_jt_i}}{\sum\limits_{j=1}^{P}\alpha_j}\right)\right.\\\left.+\sum\limits_{k=1}^{i-1}\sum\limits_{j=1}^{P=2}\alpha_j\mathrm{e}^{-\beta_j(t_i-t_k)}\right].
\end{aligned}
\end{equation}

\section{Asymmetric multivariate case}
\label{appendix:asymmetric}

The mutual influence in the asymmetric case is defined as

\begin{displaymath} \boldsymbol{\alpha} = \left( \begin{array}{ccc} \alpha_0 & \alpha_{m}^1 \\
\alpha_{m}^2 & \alpha_0 \\
\end{array} \right), \end{displaymath}
where $\alpha_m^1\neq\alpha_m^2$ (specifically $\alpha_m^1<\alpha_m^2$ by convention here). The largest eigenvalue is now given by 
\begin{equation}
\rho(\Gamma)=\frac{\alpha_0+\sqrt{\alpha_m^1\alpha_m^2}}{\beta}.
\end{equation}

In the asymmetric case, we see that the parameter estimates of the backward arrow of time are very significantly different from those of the forward arrow of time. Even more, in about 75\% cases the backward-arrow fits produce almost nonsensical results or simply do not converge. When these runs are removed, there is a bias towards keeping the runs which are significantly longer than the desired number of events, which might help explain the misestimation of certain parameters.

 \begin{figure}
\centering
~~~\includegraphics[width=0.9\columnwidth]{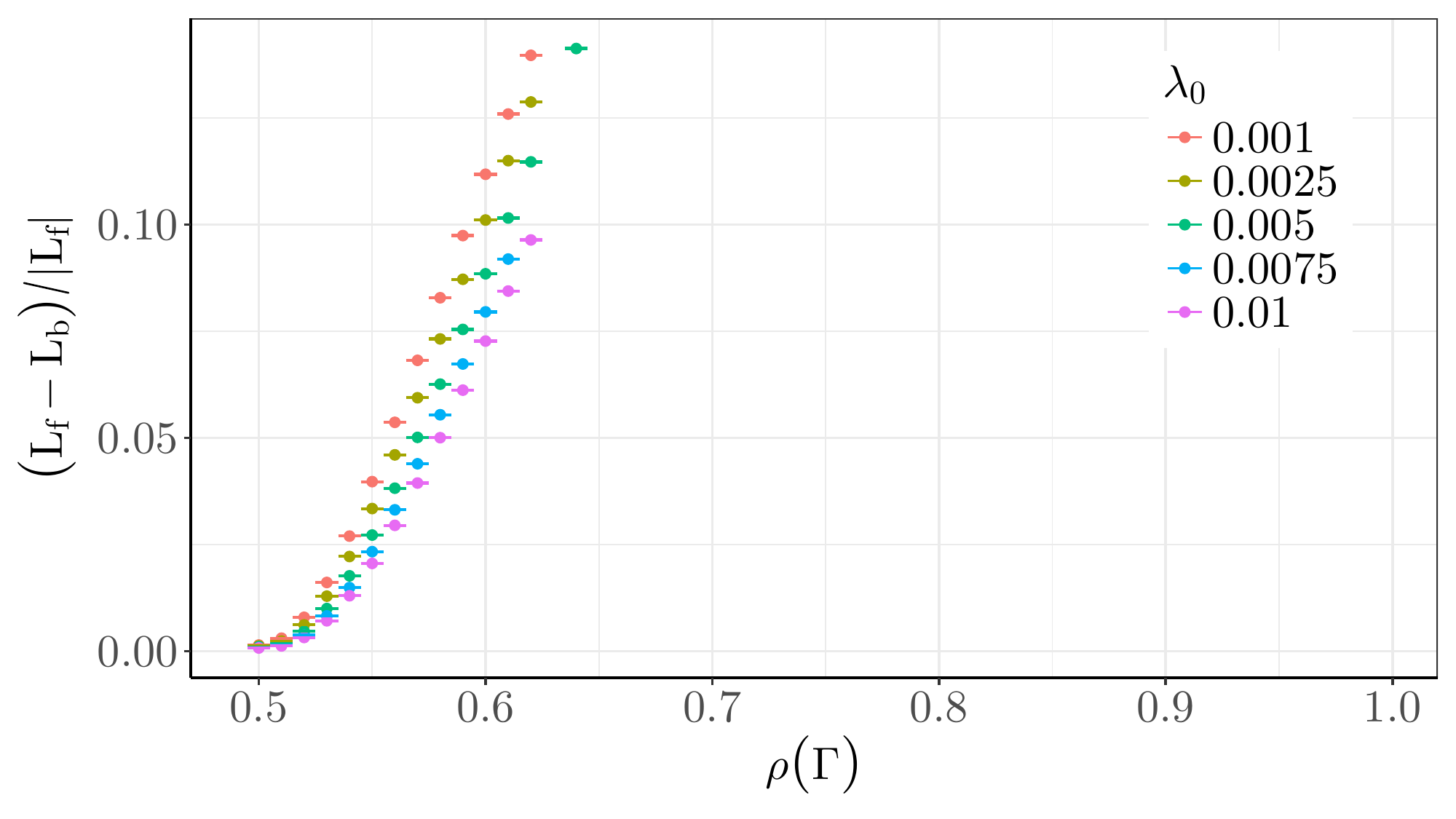}  
~~~\includegraphics[width=0.9\columnwidth]{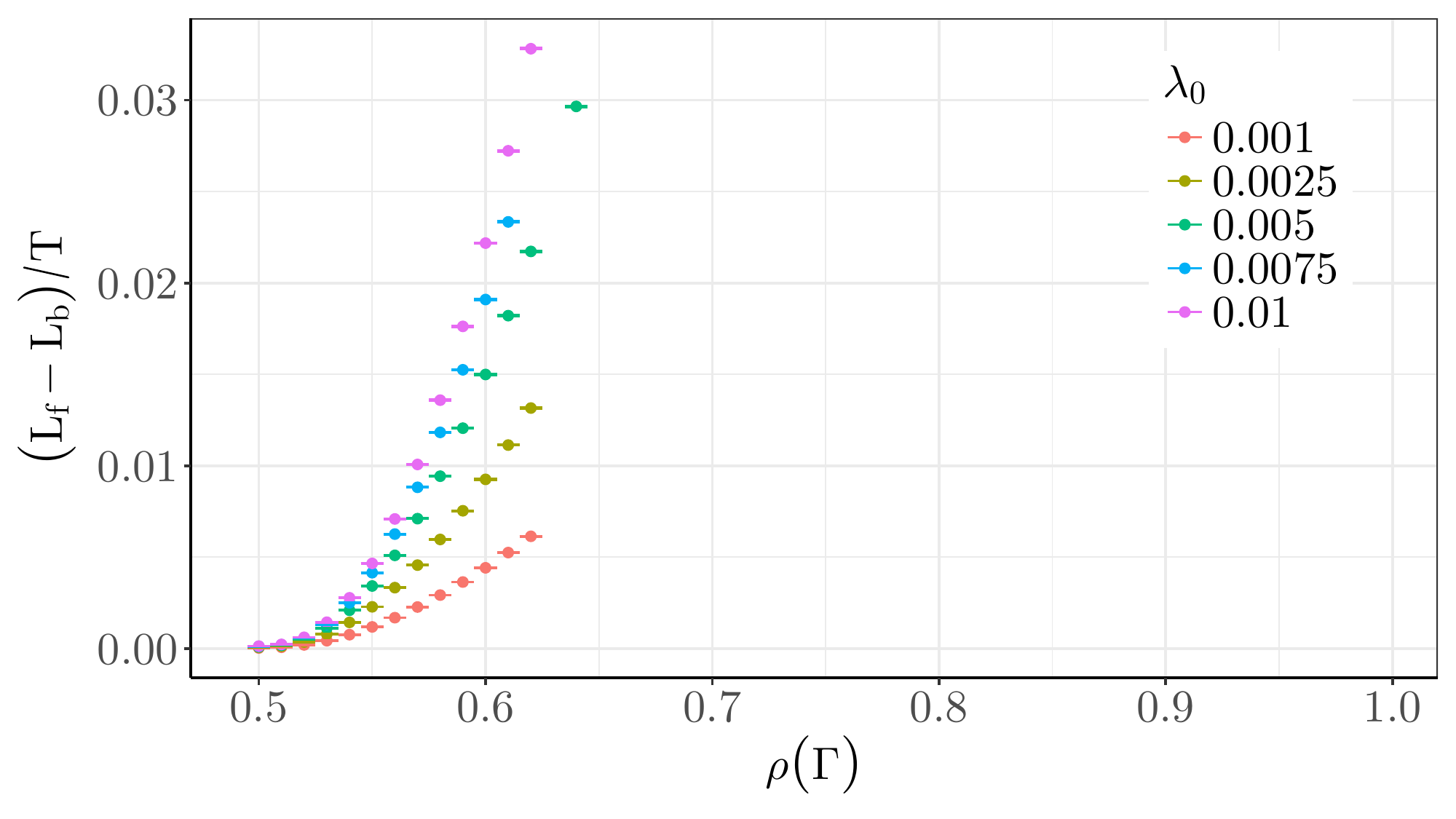} 

\caption{Relative difference of the log-likelihood between forward and backward time arrows (top) and difference of the log-likelihood between forward and backward time arrows with regards to $T$ (bottom) for a multidimensional HP with an asymmetric excitation kernel. All possible permutations of $\lambda_0=\{0.0010, 0.0025, 0.0050, 0.0075, 0.100\}$, $\alpha_m^1=\{0.049\}$, with $\alpha_m^2$ chosen according to the desired maximum eigenvalue $\rho(\Gamma)$ and $\alpha_m^1<\alpha_m^2$, and $\beta=0.1$ are considered. The data points are grouped according to maximum eigenvalue and averaged over 100 runs for each parameter permutation. The expected total number of events is set to $10^6$.}
\label{fig:loglik_reldiff_mHP_asymm}
\end{figure}

\begin{figure}
\centering
\includegraphics[width=0.9\columnwidth]{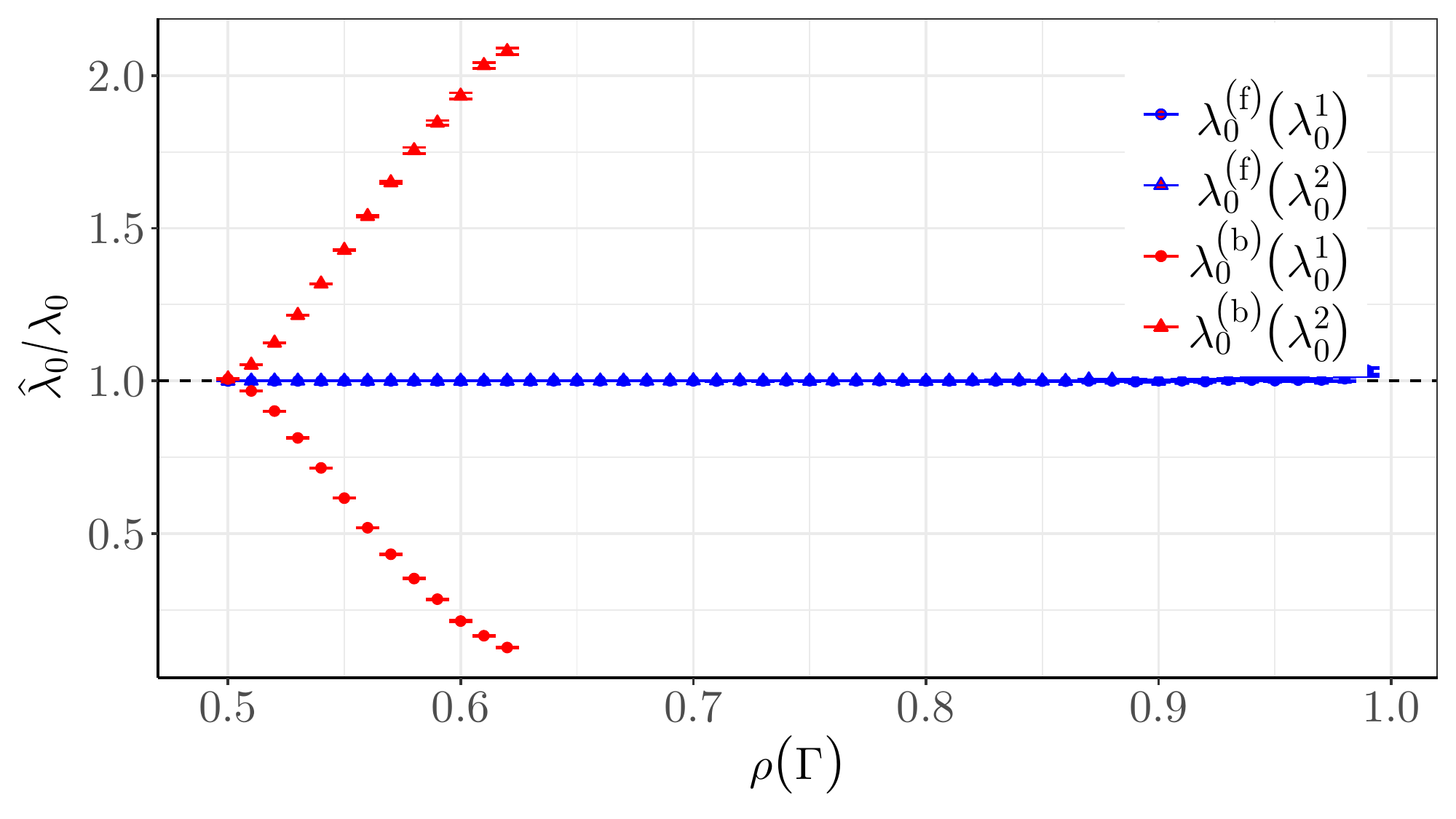}
\includegraphics[width=0.9\columnwidth]{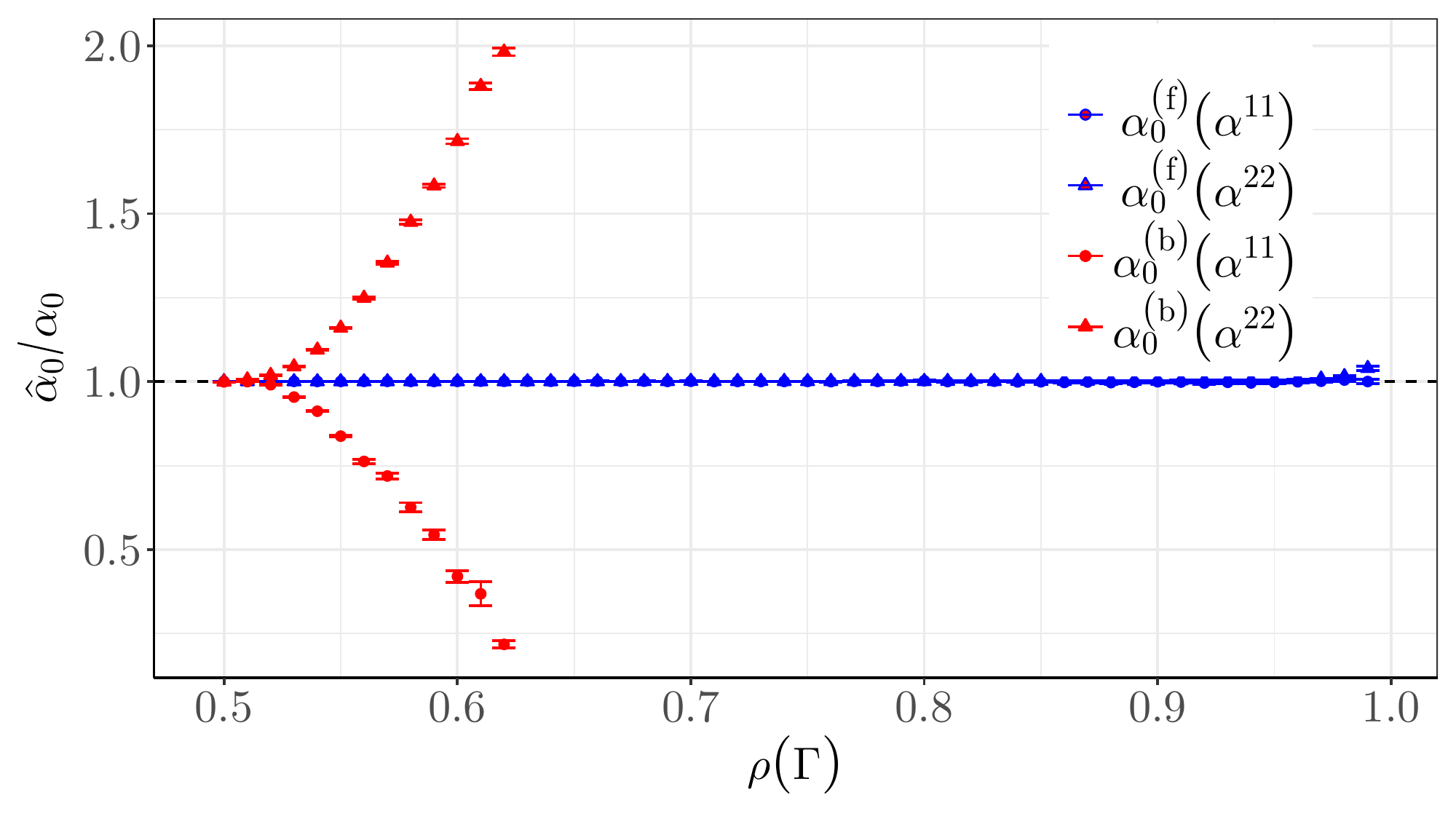}
\includegraphics[width=0.9\columnwidth]{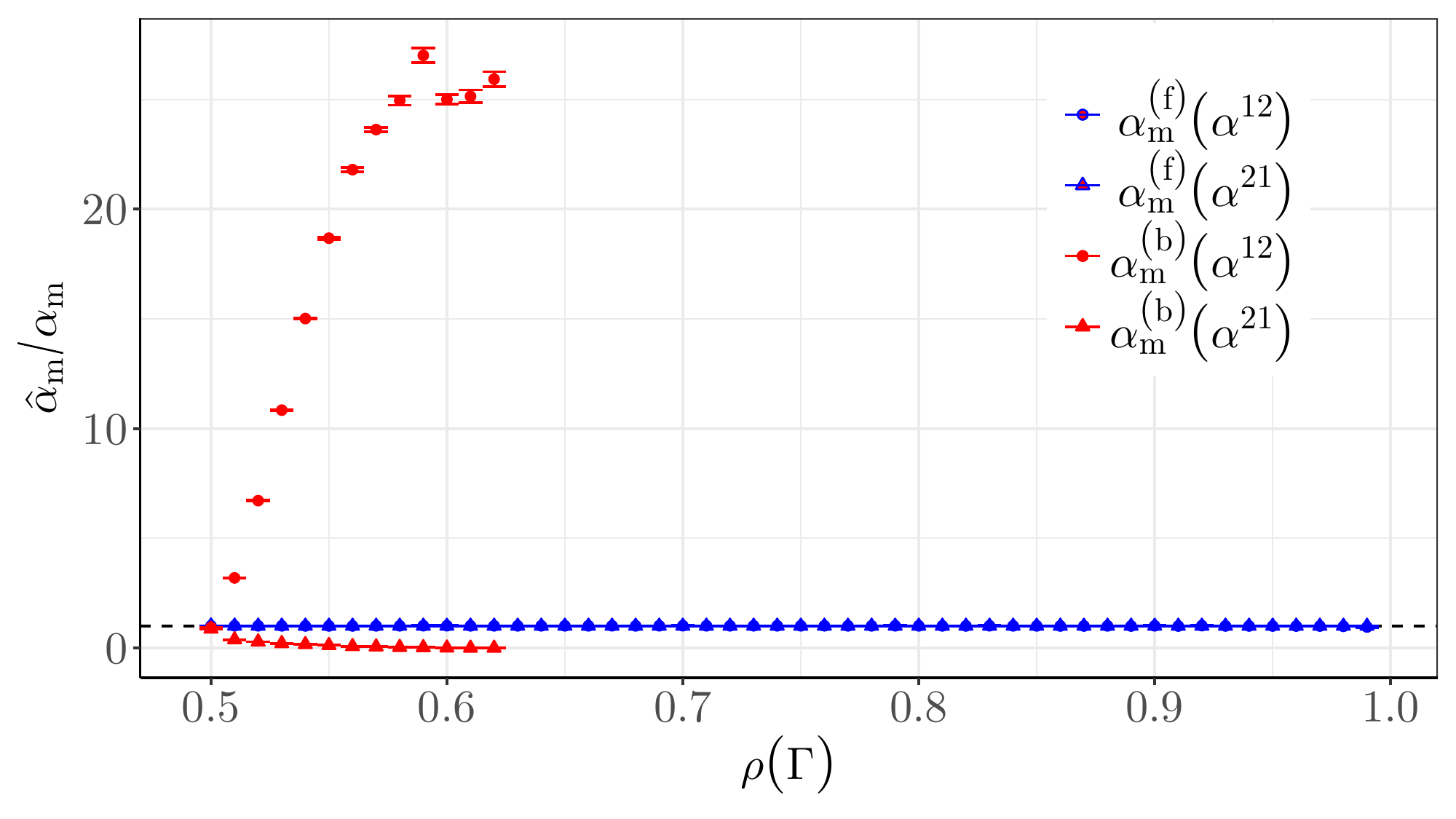}
\includegraphics[width=0.9\columnwidth]{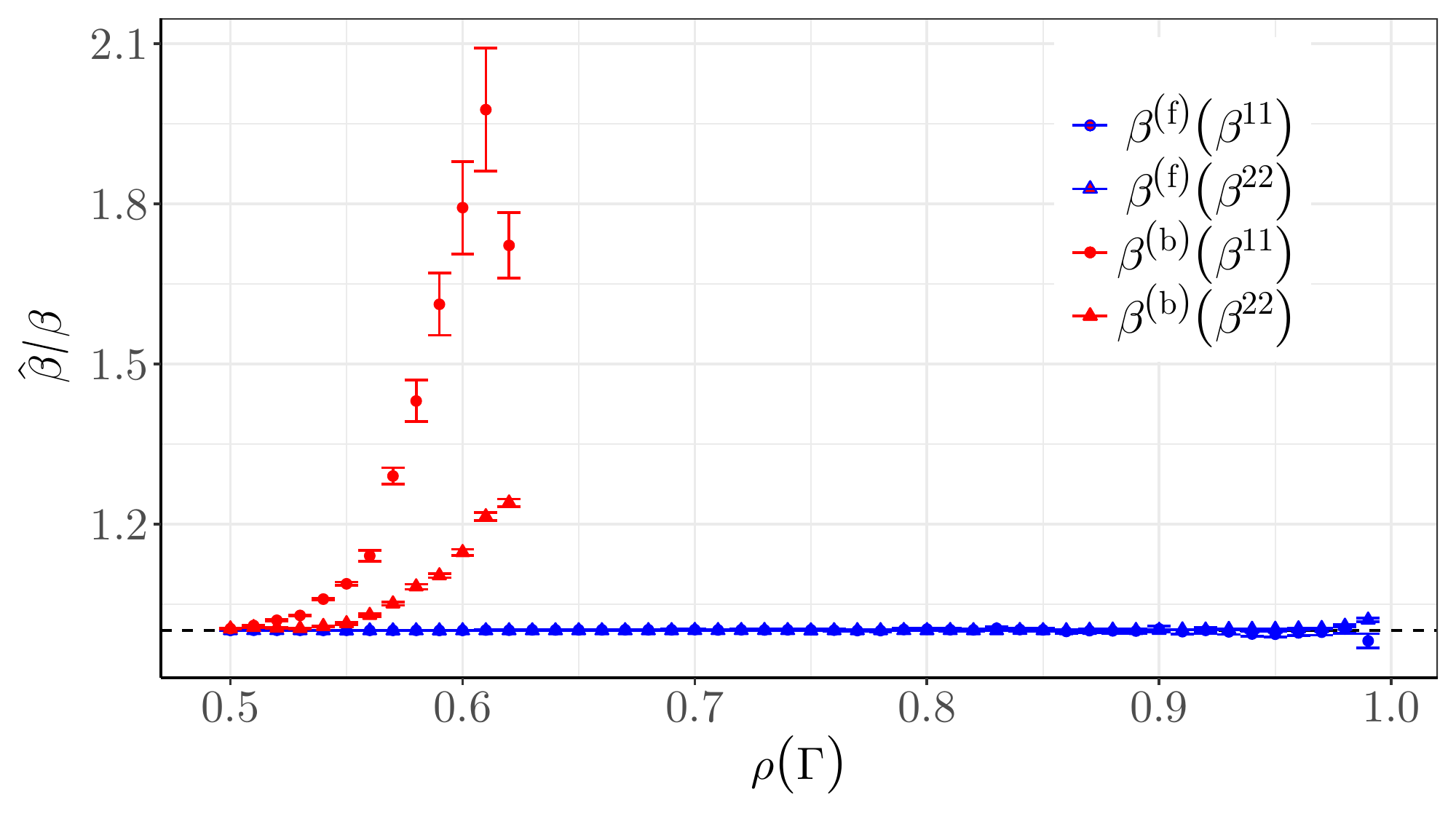}
\caption{Relative difference in the estimation of the various parameters in the MLE of the multidimensional HP with an asymmetric excitation matrix for the forward (blue) and the backward process (red). All possible permutations of $\lambda_0=\{0.0010, 0.0025, 0.0050, 0.0075, 0.100\}$, $\alpha_m^1=\{0.049\}$, with $\alpha_m^2$ chosen according to the desired maximum eigenvalue $\rho(\Gamma)$ and $\alpha_m^1<\alpha_m^2$, and $\beta=0.1$ are considered. The data points are grouped according to maximum eigenvalue and averaged over 100 runs for each parameter permutation. The expected total number of events is set to $10^6$.}
\label{fig:fig_par_mHP_asymm}
\end{figure}

\begin{figure}
\centering
\includegraphics[width=0.9\columnwidth]{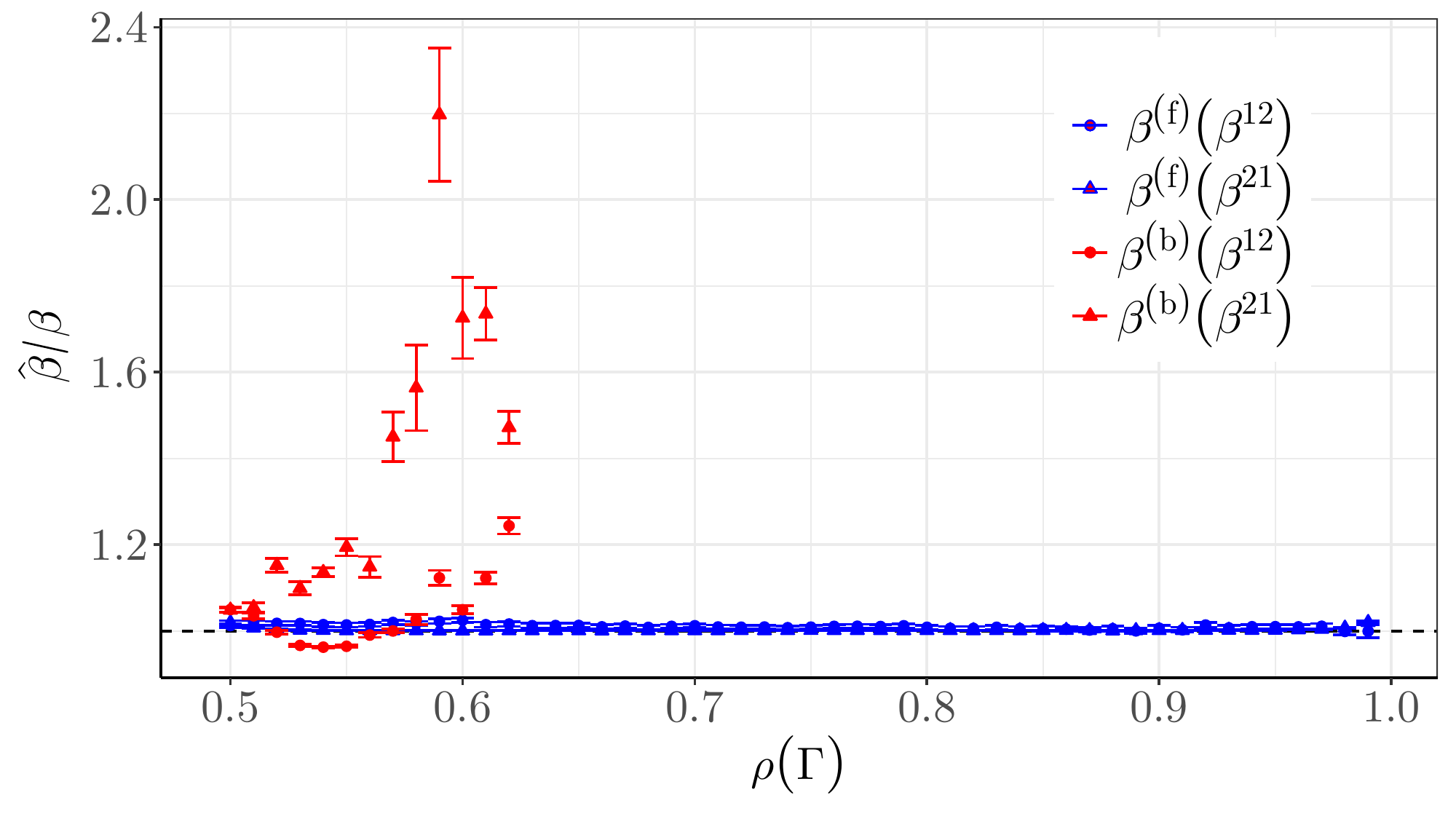}
\includegraphics[width=0.9\columnwidth]{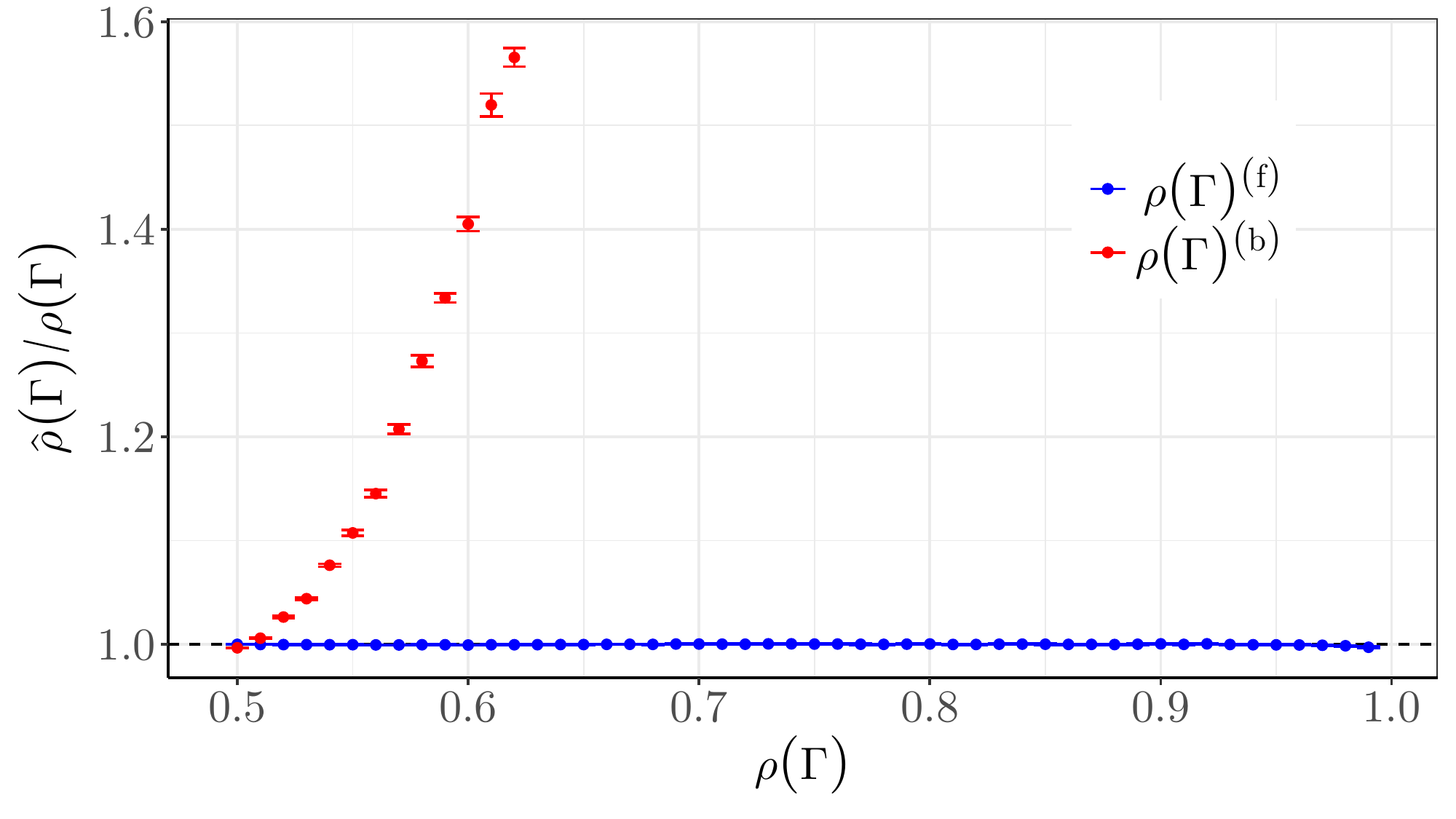}
\caption{ (Continued) Relative difference in the estimation of the various parameters in the MLE of the multidimensional HP with an asymmetric excitation matrix for the forward (blue) and the backward process (red). All possible permutations of $\lambda_0=\{0.0010, 0.0025, 0.0050, 0.0075, 0.100\}$, $\alpha_m^1=\{0.049\}$, with $\alpha_m^2$ chosen according to the desired maximum eigenvalue $\rho(\Gamma)$ and $\alpha_m^1<\alpha_m^2$, and $\beta=0.1$ are considered. The data points are grouped according to maximum eigenvalue and averaged over 100 runs for each parameter permutation. The expected total number of events is set to $10^6$.}
\label{fig:fig_par_mHP_asymm_cont}
\end{figure}

\end{document}